\documentclass[pra,twocolumn,floatfix,showpacs,nofootinbib]{revtex4}
\usepackage{hypernat}

\usepackage{hyperref,rotating,enumerate,longtable,dcolumn,graphicx,graphics,times,amsmath,amsfonts,amssymb,latexsym,bm,bbm,textcomp,dsfont,natbib}

\newcommand{\ignore}[1]{}
\begin{document}

\title{Intrinsic geometry of quantum adiabatic evolution and quantum phase transitions}
\author{A. T. Rezakhani$^{(1,4)}$, D. F. Abasto$^{(2,4)}$, D. A. Lidar$^{(1,2,3,4)}$, and P. Zanardi$^{(2,4,5)}$}
\affiliation{$^{(1)}$Departments of Chemistry, $^{(2)} $Physics, and
  $^{(3)}$Electrical Engineering, and $^{(4)}$Center for Quantum Information Science \&
Technology, University of Southern California, Los Angeles, California 90089, USA\\
$^{(5)}$Institute for Scientific Interchange, Viale Settimio Severo 65, I-10133 Torino, Italy
}

\begin{abstract}
We elucidate the geometry of quantum adiabatic
evolution. By minimizing the deviation
from adiabaticity we find a Riemannian metric tensor
underlying adiabatic evolution. Equipped with this tensor, we identify a unified geometric description
of quantum adiabatic evolution and quantum phase transitions, which generalizes previous treatments to
allow for degeneracy. The same structure is relevant for applications
in quantum information processing, including adiabatic and holonomic
quantum computing, where geodesics over the manifold of control
parameters correspond to paths which minimize errors. We illustrate this geometric structure with
examples, for which we explicitly find adiabatic geodesics. By
solving the geodesic equations in the vicinity of a quantum critical
point, we identify universal characteristics of optimal adiabatic
passage through a quantum phase transition. In particular, we show
that in the vicinity of a critical point describing a second order
quantum phase transition, the geodesic exhibits
power-law scaling with an exponent given by twice the inverse of the product
of the spatial and scaling dimensions.
\end{abstract}

\maketitle

%%%%%%%%%%%%%%%%%%%%%%%%%%%%%%%%%%%%%%%%%%%%%%%%%%%%%%%%%%
\section{Introduction}

Geometric and topological concepts have long played useful roles in both
classical and quantum physics \cite{Nakahara-book}. Important applications
where the use of geometry has led to new insights include quantum
evolutions \cite{AnandanAharonov:90}, distance measures in quantum
information theory \cite{Information1,Information2}, circuit-based
quantum computation \cite{Nielsen-science}, and 
holonomic quantum computation \cite{HQC}. More recently quantum phase
transitions (QPTs) \cite{Sachdev-book} and adiabatic quantum computation \cite{Farhi1,Farhi2} have also been been explored from a geometric perspective 
\cite{Zanardi-prl:07,QAB}. While geometry can be seen as an underlying
unifying theme in these applications, an explicit geometry-based
connection between them is not always apparent. The central theme of this work is to elucidate the geometry of
adiabatic evolution. In particular, we describe an
all-geometric connection between QPTs and adiabatic quantum evolution. We do this
by showing how the Riemannian metric tensor that describes
transitions through quantum critical points \cite{Zanardi-prl:07} also
arises in adiabatic quantum evolution.
More specifically, we explain how the metric which provides an
information-geometric framework for QPTs can also provide a geometry for the
control manifold arising in adiabatic evolutions. That QPTs and
adiabatic quantum evolution should be so intimately related
was previously understood in terms of the role of
ground state evolution in adiabatic quantum computation,
and in particular the basic observation that those points where ground state properties undergo drastic changes, i.e.,
quantum critical points, are bottlenecks for adiabaticity \cite{Farhi1,Latorre,Amin}.

The metric tensor we identify is a natural extension of the metric found in
Ref.~\cite{Zanardi-prl:07} to systems with degenerate ground states. In this sense we go beyond
adiabatic quantum computation, which is typically concerned with nondegenerate ground states, and find
results with applications to holonomic quantum computation, where quantum gates are performed as
holonomies in the degenerate ground eigensubspace of the system Hamiltonian.
We analyze the relevance of the metric tensor we identify for determining
paths with minimum computational error, in the sense of deviation from the
desired final adiabatic state. In addition, we find a prescription for
adiabatic passage through quantum critical regions by solving the
corresponding geodesic equations derived from the metric tensor.
As a result we are able to identify universal characteristics of
adiabatic passage through a critical point. Namely, we find that in the vicinity of a critical point the geodesic exhibits
power-law scaling with an exponent given by twice the inverse of the product
of the spatial and scaling dimensions.

The structure of this paper is as follows.  In Sec.~\ref{secII} we
formulate our geometric picture. Specifically, after defining the model in subsection \ref{model},
in subsection \ref{Ad-Er} we introduce the adiabatic error and show how to upper
bound it as a sum of two components, one of which
encodes the geometric aspects of the evolution. We obtain a Riemannian
metric by minimizing this error. Next, in subsection~\ref{Op-Fi} we demonstrate
the emergence of the same geometry from the concept of adiabatic operator
fidelity. In subsection \ref{Nat} we demonstrate
how our metric arises from
three
more (interrelated) natural origins: Grassmannian geometry, Uhlmann parallel
transport,
and the Bures metric.
In subsection \ref{Comp} we compare
our metric with another, related metric for adiabatic evolutions
which we proposed in earlier work \cite{QAB}.
We briefly discuss strategies for further making the
adiabatic error small in subsection \ref{Red}.
We make the connection to QPTs in section \ref{QPT}.
Specifically, in subsection~\ref{QPT-metricg} we establish the
relevance of our metric in the sense of QPTs, by showing that the same
metric is responsible for signaling quantum criticality. Then, in
subsection \ref{g-criticality} we derive the quantum critical
scaling of the metric tensor. Switching gears, we
define the notion of an adiabatic geodesic in Sec.~\ref{secIII}.
In subsection \ref{examples} we analyze three
examples, namely the Deutsch-Jozsa algorithm, projective Hamiltonians
(including Grover's algorithm), and the transverse field Ising model,
for which we analytically find the adiabatic metric and the corresponding geodesics.
In subsection \ref{geodesic} we analyze the properties of geodesics
when the adiabatic evolution passes through a quantum critical
point. It is here that we identify the universal characteristics of
such geodesics. We summarize our results and conclude
in Sec.~\ref{Summ}. Several appendices provide detailed proofs omitted from the main text
so as not to interrupt the presentation.

%%%%%%%%%%%%%%%%%%%%%%%%%%%%%%%%%%%%%%%%%%%%%%%%%%%%%%%%%%%
\section{Geometry of adiabatic quantum evolution}
\label{secII}

\subsection{Model}
\label{model}

Consider an $n$-body system with the $N$-dimensional Hilbert space $\mathcal{%
H}$. The Hamiltonian family $\{H(\mathbf{x})\}$ for this system, which
depends on the (time-dependent) coupling strengths or \textquotedblleft
control knobs\textquotedblright\ $\mathbf{x}$, can be identified by points
over the real $M$-dimensional manifold $\mathcal{M}\ni \mathbf{x}$. Given a
total evolution time $T$ and rescaled time $s=t/T$, a path $\mathbf{x}:s\in
\lbrack 0,1]\mapsto \mathcal{M}$ then represents the dynamics in this time
interval, starting from $\mathbf{x}_{0}\equiv \mathbf{x}(0)$ and ending at $%
\mathbf{x}_{1}\equiv \mathbf{x}(1)$. We shall use the notation $\mathbf{x}%
_{s}\equiv \mathbf{x}(s)$ interchangeably, or sometimes drop the $s$%
-dependence entirely to lighten the notation. We allow for a $g_{0}(\mathbf{x%
})$-fold degenerate ground-state eigensubspace of $\{H(\mathbf{x})\}$, with
eigenstates $\{|\Phi _{0}^{\alpha }(\mathbf{x})\rangle \}$. Thus this
subspace can be identified by the projector 
\begin{equation}
P_{0}(\mathbf{x})=\sum_{\alpha =1}^{g_{0}(\mathbf{x})}|\Phi _{0}^{\alpha }(%
\mathbf{x})\rangle \langle \Phi _{0}^{\alpha }(\mathbf{x})|,
\end{equation}%
with $\mathrm{Tr}[P_{0}(\mathbf{x})]=g_{0}(\mathbf{x})\geq 1$. We assume
that for all finite $n$ the ground-state energy $E_{0}(\mathbf{x})$ is
separated by a nonvanishing gap $\Delta (\mathbf{x})$ from the rest of the
spectrum. In the thermodynamic limit $n\rightarrow \infty $ we allow the gap
to vanish at some finite set of points $\{\mathbf{x}_{c}\equiv \mathbf{x}%
(s_{c})\}$, or a segment of the path. These are the critical points where a
QPT\ takes place. Although our results would hold if we picked any other
eigensubspace satisfying the previous requirements, rather than the ground
state, for specificity we shall henceforth consider the ground state and the
initialization $|\psi (0)\rangle =\sum_{\alpha =1}^{g_{0}(\mathbf{x}%
_{0})}a_{\alpha }|\Phi _{0}^{\alpha }(\mathbf{x}_{0})\rangle $ (where $|\psi
(s)\rangle \equiv |\psi \left( \mathbf{x}_{s}\right) \rangle $, and we
similarly drop the explicit dependence on $\mathbf{x}(s)$ hereafter where
possible).

%%%%%%%%%%%%%%%%%%%%%%%%%%%%%%%%%%%%%%%%%%%%%%%%%%%%%%%%%%%
\subsection{Adiabatic Error}
\label{Ad-Er}

\subsubsection{Degenerate case}

We wish to compare the desired, \textquotedblleft ideal\textquotedblright\
adiabatic evolution to the actual evolution induced by the the Hamiltonian
family. To this end we shall define an appropriate \textquotedblleft
adiabatic error\textquotedblright\ which measures the deviation between the
two. The state of the system, 
\begin{equation}
|\psi (s)\rangle =V(s)|\psi (0)\rangle ,
\end{equation}%
at any rescaled time $s$, is given in $\hbar =1$ units [adopted hereafter], in terms of the
propagator $V(s)$ which is the solution to the time-dependent Schr\"{o}%
dinger equation 
\begin{equation}
i\partial _{s}V(s)=TH(s)V(s).
\label{V-eq}
\end{equation}%
We can similarly associate an adiabatic propagator $V_{\text{ad}}(s)$ and an
adiabatic Hamiltonian $H_{\text{ad}}(s)$ to the ideal adiabatic evolution,
where the two are related via the Schr\"{o}dinger equation 
\begin{equation}
i\partial _{s}V_{\text{ad}}(s)=TH_{\text{ad}}(s)V_{\text{ad}}(s).
\label{Vad-eq}
\end{equation}%
What defines the adiabatic propagator is the \textquotedblleft intertwining
property\textquotedblright\ 
\begin{equation}
V_{\text{ad}}(s)P_{0}(0)V_{\text{ad}}^{\dag }(s)=P_{0}(s),
\label{intertw}
\end{equation}%
which means that $V_{\text{ad}}(s)$ preserves the band structure of the
ground eigensubspace of $H(s)$. By differentiation the intertwining property
is equivalent to $i\partial _{s}P_{0}(s)=T[H_{\text{ad}}(s),P_{0}(s)]$, and
when it holds we have%
\begin{equation}
|\psi _{\text{ad}}(s)\rangle =V_{\text{ad}}(s)|\psi (0)\rangle =\sum_{{%
\alpha \alpha ^{\prime }=0}}^{g_{0}}a_{\alpha }V_{\alpha \alpha ^{\prime
}}^{[0]}(s)|\Phi _{0}^{\alpha ^{\prime }}(s)\rangle ,
\end{equation}%
where $V_{\alpha \alpha ^{\prime }}^{[0]}(s)=\langle \Phi _{0}^{\alpha }(s)|V_{\text{ad}}(s)|\Phi _{0}^{\alpha^{\prime
} }(0)\rangle $ is the (non-Abelian)
Wilczek-Zee holonomy \cite{WZ:84}---usually expressed as the path-ordered
exponential 
\begin{equation}
V^{[0]}(s)=\mathcal{P}\exp\Big(- \int_{0}^{s}A(s^{\prime })\mathrm{d}s^{\prime }\Big),
\label{WZFORMULA}
\end{equation}%
with the gauge connection 
\begin{equation}
A_{\alpha \alpha ^{\prime }}\equiv \langle \Phi _{0}^{\alpha }|\partial
_{s}|\Phi _{0}^{\alpha ^{\prime }}\rangle .
\end{equation}
We prove Eq.~(\ref{WZFORMULA}) in Appendix~\ref{app:WZformula-proof} (see also Ref.~\cite{Vidal:91}).

The adiabatic Hamiltonian can be expressed in terms of the original
Hamiltonian plus a \textquotedblleft correction\textquotedblright\ term \cite{Avron:871,Avron:872}:%
\begin{equation}
H_{\text{ad}}(s)=H(s)+i[\partial _{s}P_{0}(s),P_{0}(s)]/T,  \label{Had-H}
\end{equation}%
Clearly then, 
the actual state $|\psi (s)\rangle $ need not be the same as
the adiabatic state $|\psi _{\text{ad}}(s)\rangle $. Our objective is to
find the path $\mathbf{x}_{s}$ that minimizes the adiabatic error $\Vert
|\psi (\mathbf{x}_{s})\rangle -|\psi _{\text{ad}}(\mathbf{x}_{s})\rangle
\Vert =\Vert \{V(\mathbf{x}_{s})-V_{\text{ad}}(\mathbf{x}_{s})\}|\psi (%
\mathbf{x}_{0})\rangle \Vert $, where the norm is the standard Euclidean
norm: $\Vert |\phi \rangle \Vert \equiv \sqrt{\langle \phi |\phi \rangle }$.
However, so as to obtain a result which does not depend on the initial state 
$|\psi (\mathbf{x}_{0})\rangle $ we shall adopt a state-independent error
measure, and define the adiabatic error to be 
\begin{equation}
\delta \lbrack \mathbf{x}(s)]\equiv \Vert V(\mathbf{x}_{s})-V_{\text{ad}}(%
\mathbf{x}_{s})\Vert .  \label{del-def}
\end{equation}%
Since $\Vert (V-V_{\text{ad}})|\psi \rangle \Vert \leq \Vert V-V_{\text{ad}%
}\Vert $, where the norm on the right-hand side is the standard sup-operator
norm (often denoted $\Vert \cdot \Vert _{\infty }$) \cite{Bhatia:book} 
\begin{equation}
\Vert X\Vert \equiv \sup_{|v\rangle :~\Vert |v\rangle\Vert =1}\sqrt{%
\langle v|X^{\dag }X|v\rangle }=\max_{i}\sigma _{i}(X),
\end{equation}
where $\{\sigma _{i}(X)\}$ are the singular values of $X$ (eigenvalues of $%
\sqrt{X^{\dag }X}$), an upper bound on $\delta \lbrack \mathbf{x}(s)]$ is
then also an upper bound on $\Vert |\psi (\mathbf{x}_{s})\rangle -|\psi _{%
\text{ad}}(\mathbf{x}_{s})\rangle \Vert $.

Using the fact that the sup-operator norm is unitarily invariant ($\Vert
VAW\Vert =\Vert A\Vert $ for any operator $A$ and any pair of unitaries $V$
and $W$) we can rewrite $\delta $ as 
\begin{equation}
\delta \lbrack \mathbf{x}(s)]=\Vert I-\Omega (\mathbf{x}_{s})\Vert ,
\end{equation}%
where the \textquotedblleft wave operator\textquotedblright\ 
\begin{equation}
\Omega (s)\equiv V_{\text{ad}}^{\dag }(s)V(s)  \label{Om-def}
\end{equation}%
satisfies the Volterra equation 
\begin{equation}
\Omega (s)=I-\int_{0}^{s}K_{T}(s^{\prime })\Omega (s^{\prime })\mathrm{d}%
s^{\prime },  \label{Om}
\end{equation}%
with the kernel 
\begin{equation}
K_{T}(s)\equiv V_{\text{ad}}^{\dag }(s)[\partial _{s}P_{0}(s),P_{0}(s)]V_{\text{ad}}(s).
\label{K-def}
\end{equation}%
Considering Eq.~(\ref{Had-H}), 
$-iK_{T}(s)/T$
is simply the
interaction-picture Hamiltonian which results from transforming $H(s)$ to
the interaction picture with respect to $H_{\text{ad}}(s)$, where $%
i[\partial _{s}P_{0}(s),P_{0}(s)]/T$ plays the role of the perturbation.
Therefore, in analogy to the Dyson series of time-dependent perturbation
theory, the Volterra equation can be solved by iteration, which yields 
\begin{equation}
\Omega (s)=\sum_{l=0}^{\infty }\Omega _{l}(s),
\end{equation}%
where 
\begin{eqnarray}
\Omega _{0}(s) &=&I, \\
\Omega _{l>0}(s) &=&-\int_{0}^{s}K_{T}(s^{\prime })\Omega _{l-1}(s^{\prime })%
\mathrm{d}s^{\prime }.
\label{omega_l}
\end{eqnarray}%
As shown in Refs.~\cite{Avron:871,Avron:872}, $\forall l\in \{2k-1,2k\}$
$(k\in \mathds{N})$: 
\begin{eqnarray}
&& \sup_{s}\Vert \Omega _{l}(s)\Vert =\mathcal{O}(1/T^{k}),  \label{Avron1} \\
&& \sup_{s}\Vert \Omega (s)-\sum_{j=0}^{l-1}\Omega _{j}(s)\Vert =\mathcal{O}%
(1/T^{k}).  \label{Avron2}
\end{eqnarray}%
Using the above results, $\Vert I-\Omega (s)\Vert $ can be expressed in terms of
a $1/T$ series expansion, since 
\begin{eqnarray}
\Vert I-\Omega (s)\Vert &=&\Vert \Omega _{1}(s)-\sum_{l\geq
2}\int_{0}^{s}K_{T}(s^{\prime })\Omega _{l-1}(s^{\prime })\mathrm{d}%
s^{\prime }\Vert  \notag \\
&\leq &\Vert \Omega _{1}(s)\Vert + \notag \\
&&\int_{0}^{s}\Vert K_{T}(s^{\prime })\Vert
\sum_{l\geq 2}\Vert \Omega _{l-1}(s^{\prime })\Vert \mathrm{d}s^{\prime }
\label{I-Om1} \\
&=&\Vert \Omega _{1}(s)\Vert +\widetilde{\epsilon}(s)\mathcal{O}(1/T),
\label{I-Om2}
\end{eqnarray}%
where
\begin{equation}
\widetilde{\epsilon}(s)\equiv \int_{0}^{s}\Vert \lbrack \partial _{s^{\prime
}}P_{0}(s^{\prime }),P_{0}(s^{\prime })]\Vert \mathrm{d}s^{\prime }.
\label{eps-def}
\end{equation}%
Thus the error $\delta $ is upper-bounded as 
\begin{equation}
\delta \lbrack \mathbf{x}(s)]\leq \delta _{1}(s)+\delta _{2}[\mathbf{x}(s)],
\label{important}
\end{equation}%
where%
\begin{eqnarray}
\delta _{1}(s) &\equiv &\Vert \Omega _{1}(s)\Vert \sim \mathcal{O}(1/T)
\label{del1} \\
\delta _{2}[\mathbf{x}(s)] &\equiv &\widetilde{\epsilon}[\mathbf{x}(s)]\mathcal{O%
}(1/T).  \label{del2}
\end{eqnarray}%
Both error components can evidently be made small by choosing a large $T$,
while for a given $T$, $\delta _{2}$ can additionally be made small by
choosing a path over the control manifold $\mathcal{M}$ with small $\widetilde{%
\epsilon}$. Note that in addition to $\Vert \Omega _{1}(s)\Vert \sim 
\mathcal{O}(1/T)$ we also have the bound $\Vert \Omega _{1}(s)\Vert \leq
\int_{0}^{s}\Vert K_{T}(s^{\prime })\Vert \mathrm{d}s^{\prime }=\widetilde{%
\epsilon}[\mathbf{x}(s)]$, but from Eqs.~(\ref{Avron1}) and
(\ref{Avron2}) as such we do not have a bound of the form $% 
\Vert \Omega _{1}(s)\Vert \leq \widetilde{\epsilon}[\mathbf{x}(s)]\mathcal{O}%
(1/T)$. 
One can see from Ref.~\cite{JRS}
how $\delta_1(s)$ depends on $T$, the gap, and the norm of the
Hamiltonian or its derivatives.
However, the
coefficient of the $1/T$ term of $\delta_1$
does not appear to have a geometric significance, and we shall therefore
exclude $\delta_1$ from our study of adiabatic geometry in this paper.

In the following we shall make the upper bound on
$\delta$ small by finding a path which
makes $\widetilde{\epsilon}[\mathbf{x}(s)]$ small.
Finding the path which minimizes $\widetilde{\epsilon}$ is, however, beyond the scope of
this work. Instead, as we show below, after
replacing the sup-operator norm by the Frobenius norm, the problem of minimizing $\delta
_{2}$ has a geometric solution in the sense that a Riemannian metric tensor is encapsulated in $\epsilon \lbrack \mathbf{x}(s)]$ [Eq.~(\ref{eps-def}) with the modified norm].

In Appendix~\ref{app:PdotP-norm-proof} we prove that 
\begin{equation}
\Vert \lbrack \partial _{s}P_{0},P_{0}]\Vert =\sqrt{\Vert P_{0}(\partial
_{s}H)\left( \frac{1}{H-E_{0}}\right) ^{2}(\partial _{s}H)P_{0}\Vert },
\label{PDOTP-NORM}
\end{equation}
where $[H-E_0]^{-1}$ is called the reduced resolvent and is a
shorthand for $(I-P_0)[H-E_0]^{-1}(I-P_0)$.

For a different method of traversing eigenstate paths of Hamiltonians,
based on the use of evolution randomization and a quantum phase estimation
algorithm, see Ref.~\cite{Boixo}.

%%%%%%%%%%%%%%%%%%%%%%%%%%%%%%%%%%%%%%%%%%%%%%%%%%%%%%%%%%%
\subsubsection{Nondegenerate case}

When $H$ has a discrete and nondegenerate spectrum, $P_{0}=|\Phi _{0}\rangle
\langle \Phi _{0}|$ and $I-P_{0}=\sum_{n>0}|\Phi _{n}\rangle \langle \Phi
_{n}|$, where $\{|\Phi _{n}\rangle \}_{n>0}$ are the excited eigenstates of $%
H$ with eigenvalues $\{E_{n}\}_{n>0}$. In this case 
\begin{equation}
\frac{1}{H-E_{0}}=\sum_{n>0}\frac{1}{E_{n}-E_{0}}|\Phi _{n}\rangle \langle
\Phi _{n}|.  \label{resolvent}
\end{equation}%
Using the chain rule of differentiation to write $\partial _{s}H=(\partial
_{i}H)\dot{\mathbf{x}}^{i}$, where dot denotes $\partial _{s}$ and $\partial
_{i}$ denotes $\partial /\partial \mathbf{x}^{i}$, and using the Einstein summation convention
Eq.~(\ref{PDOTP-NORM}) is easily simplified in the nondegenerate
case to yield:%
\begin{equation}
\widetilde{\epsilon}[\mathbf{x}(s)]=\int_{0}^{s}\sqrt{2\mathbf{g}_{ij}^{(1)}(%
\mathbf{x})\dot{\mathbf{x}}^{i}\dot{\mathbf{x}}^{j}}\mathrm{d}s^{\prime },
\label{eps}
\end{equation}%
where 
\begin{equation}
\mathbf{g}_{ij}^{(1)}\equiv \mathrm{Re}\left[\sum_{n>0}\dfrac{\langle \Phi _{0}|\partial
_{i}H|\Phi _{n}\rangle \langle \Phi _{n}|\partial _{j}H|\Phi _{0}\rangle }{%
(E_{n}-E_{0})^{2}}\right].  \label{metricg}
\end{equation}%
The manner in which $\mathbf{g}_{ij}^{(1)}$ appears in Eq.~(\ref{eps})
suggests that it plays the role of a metric tensor. This metric tensor is
identical to the metric tensor which was identified in the
differential-geometric theory of QPTs \cite{Zanardi-prl:07}. We next
consider how to generalize this result to the degenerate case.

%%%%%%%%%%%%%%%%%%%%%%%%%%%%%%%%%%%%%%%%%%%%%%%%%%%%%%%%%%%
\subsubsection{Metric tensor for the degenerate case -- moving to the
Hilbert-Schmidt norm}
\label{subsec:HS}

We would like to identify Eq.~(\ref{PDOTP-NORM}) with a metric tensor.
However, the appearance of the sup-operator norm presents a problem, since
this norm need not be differentiable. Hence we replace the operator norm
with the Frobenius (or Hilbert-Schmidt) norm 
\begin{equation}
\Vert X\Vert _{2}\equiv \sqrt{\text{Tr}[X^{\dag }X]}=\sqrt{\sum_{i=1}^{%
\mathrm{rank}(X)}\sigma _{i}^{2}(X)},
\end{equation}%
which satisfies \cite{Bhatia:book}%
\begin{equation}
\Vert X\Vert \leq \Vert X\Vert _{2}\leq \sqrt{\mathrm{rank}(X)}\Vert X\Vert .
\label{opnorm-ineq}
\end{equation}
Note that the operator $P_{0}(\partial _{s}H)\left( \frac{1}{H-E_{0}}\right)
^{2}(\partial _{s}H)P_{0}$ appearing in Eq.~(\ref{PDOTP-NORM}) has support
purely over the ground-state eigensubspace of $H$, due to the projections $P_{0}$
to the left and right. Therefore its rank is at most $g_{0}$, and as a
consequence of Eq.~(\ref{opnorm-ineq}) the replacement of the operator norm
by the Frobenius norm does not alter $\Vert \lbrack \partial
_{s}P_{0},P_{0}]\Vert $ (hence $\widetilde{\epsilon}$ or $\mathbf{g}$) for the
nondegenerate case ($g_{0}=1$), while it enables a differential geometric
bound in the degenerate case, which is at most $\sqrt{g_{0}}$ times greater
than the expression obtained with the operator norm. Additionally, and this
is our main reason for moving to the Frobenius norm, it guarantees
analyticity of the adiabatic error and the metric tensor when $H$ is
analytic.

With these considerations in mind, let us now redefine the adiabatic error
using the Frobenius norm:%
\begin{equation}
\epsilon (s)\equiv \int_{0}^{s}\Vert \lbrack \partial _{s^{\prime
}}P_{0}(s^{\prime }),P_{0}(s^{\prime })]\Vert _{2}\mathrm{d}s^{\prime }.
\label{eps2}
\end{equation}%
Then $\widetilde{\epsilon}(s)\leq \epsilon (s)\leq \sqrt{g_{0}}\widetilde{\epsilon}%
(s)$ and consequently 
\begin{equation}
\delta _{2}(s)\leq \epsilon (s)\mathcal{O}(1/T)\leq \sqrt{g_{0}}\delta
_{2}(s).
\end{equation}%
Minimization of $\epsilon (s)$ thus \textquotedblleft
squeezes\textquotedblright\ the error component $\delta _{2}$. We
show in Appendix \ref{app:Frob} that%
\begin{equation}
\epsilon (s)=\int_{0}^{s}\sqrt{2g_0\mathbf{g}_{ij}(\mathbf{x})\dot{\mathbf{x}}%
^{i}\dot{\mathbf{x}}^{j}}\mathrm{d}s^{\prime },
\end{equation}%
where the metric tensor is defined as%
\begin{eqnarray}
\mathbf{g}_{ij} &\equiv &\frac{1}{2g_0}\mathrm{Tr}[\partial _{i}P_{0} \partial _{j}P_{0}]
\label{g2} \\
&=&\frac{1}{2g_0}\mathrm{Tr}\left[ P_{0}(\partial _{i}H)\left( \frac{1}{H-E_{0}}\right)
^{2}(\partial _{j}H)P_{0}\right] +i\leftrightarrow j. \nonumber\\
 \label{metricH}
\end{eqnarray}%
It is simple to verify that $\mathbf{g}_{ij}$ reduces to $\mathbf{g}%
_{ij}^{(1)}$ in the nondegenerate case, and similarly $\epsilon (s)$
reduces to $\widetilde{\epsilon}(s)$ in this case.

Standard calculus of variations then tells us that minimization of $\epsilon
\lbrack \mathbf{x}(s)]$ is tantamount to finding the geodesic\ path which is
the solution to the following Euler-Lagrange (EL) equations: 
\begin{equation}
\ddot{\mathbf{x}}^{i}+\Gamma _{jk}^{i}\dot{\mathbf{x}}^{j}\dot{\mathbf{x}}%
^{k}=0,
\label{EL-eq}
\end{equation}%
where the connection $\Gamma $ is 
\begin{equation}
\Gamma _{jk}^{i}=\frac{1}{2}\mathbf{g}^{il}(\partial _{k}\mathbf{g}%
_{lj}+\partial _{j}\mathbf{g}_{lk}-\partial _{l}\mathbf{g}_{jk}).
\end{equation}%
We have thus endowed the control manifold $\mathcal{M}$ with a Riemannian
structure, given by the metric tensor\ $\mathbf{g}:T_{\mathcal{M}}\otimes T_{%
\mathcal{M}}\mapsto \mathds{R}$. That $\mathbf{g}$ really satisfies all the
properties required of a metric is shown in Appendix \ref{app:metric}. Other
geometric functions such as the curvature tensor $\mathbf{R}$ can be
calculated from $\mathbf{g}$ \cite{AvronAC}.

%%%%%%%%%%%%%%%%%%%%%%%%%%%%%%%%%%%%%%%%%%%%%%%%%%%%%%%%%%%
\subsection{Operator fidelity}
\label{Op-Fi}

Another approach to the adiabatic error is provided by the \textquotedblleft
operator fidelity\textquotedblright\ \cite{opfid} between $V$ and $V_{\text{%
ad}}$, 
\begin{equation}
f_{\varrho }[\mathbf{x}(s)]\equiv |\mathrm{Tr}[\Omega (\mathbf{x}%
_{s})\varrho ]|,  \label{f}
\end{equation}%
where $\varrho $ is an arbitrary density matrix of the system, which here we
take to be the totally mixed state $I/N$. The operator fidelity derives its
name from the fact that it quantifies the fidelity in the entire Hilbert
space, and unlike our previous error measures $\widetilde{\epsilon}$ and $%
\epsilon $, which involve the ground state projector $P_{0}$, is not
restricted just to ground states. However neither is the adiabatic error $%
\delta $ [Eq.~(\ref{del-def})] restricted just to ground states, and the two
are obviously closely related. In Appendix \ref{app:op-fid} we show that%
\begin{equation}
1-\frac{1}{\sqrt{N}}\epsilon \leq f_{\varrho}\leq 1,  \label{opfid-eps}
\end{equation}%
so that minimizing $\epsilon $ maximizes $f_{\varrho}$, and \textit{vice versa}.

Let $O$ be an arbitrary observable, and consider it in the rotated bases associated with the actual or adiabatic dynamics: 
\begin{eqnarray}
O(s) &\equiv &V(s)OV^{\dag }(s) \\
O_{\text{ad}}(s) &\equiv &V_{\text{ad}}(s)O V_{\text{ad}}^{\dag }(s), \label{Oad}
\end{eqnarray}
In addition to the bound (\ref{opfid-eps}) we show in Appendix \ref%
{app:op-fid} that
\begin{eqnarray}
\Vert O(s)-O_{\text{ad}}(s)\Vert &\leq &
\Vert O\Vert \left(\delta_1(s) +\delta
_{2}[\mathbf{x}(s)]\right)\nonumber\\
&& \left[ 2+\mathcal{O}(1/T) \right],
\label{O-bound}
\end{eqnarray}
which is identical to the adiabatic error bound (\ref{important}), apart
from the factor 
$\Vert O\Vert[2+\mathcal{O}(1/T)]$. 
Thus our bound of the operator distance $%
\Vert O(s)-O_{\text{ad}}(s)\Vert $ 
also has the component $\delta _{1}$ and the component $\delta _{2}$ with its apparent geometric contribution, which can be squeezed by choosing a geodesic path, as in subsection \ref{subsec:HS}.

%%%%%%%%%%%%%%%%%%%%%%%%%%%%%%%%%%%%%%%%%%%%%%%%%%%%%%%%%%%
\subsection{Natural geometric formulation}
\label{Nat}

\subsubsection{Grassmannian}
\label{grassmann}

An alternative, natural way to obtain a geometry for adiabatic evolutions
employs the Grassmannian structure of the dynamics \cite{Nakahara}. As
explained above, in the ideally adiabatic case the eigensubspaces
corresponding to the ground state and the rest of the spectrum ($P_{0}$ and $%
I-P_{0}$, respectively) do not mix; each follows its own unitary dynamics
determined by its Wilczek-Zee holonomy, hence $V_{\text{ad}}=V^{[0]}\oplus
V^{[\text{rest}]}$. This implies a Grassmannian manifold 
\begin{eqnarray}
\mathcal{G}_{N,g_{0}} &\cong &U(N)/U(g_{0})\times U(N-g_{0})  \notag \\
&\cong &\{P_{0}\in \mathcal{D} (\mathcal{H})~|~P_{0}^{2}=P_{0},~\mathrm{Tr}%
[P_{0}]=g_{0}\},
\end{eqnarray}%
where $U(k)$ is the group of $k\times k$ unitary matrices, and $\mathcal{D} (%
\mathcal{H})$ is the convex space of all density operators (positive semidefinite, unit trace matrices) 
defined over $\mathcal{H}$. A natural distance (metric) over
this space is given by \cite{dist-met,g-AQC-note}
\begin{equation}
d(P_{0},P_{0}^{\prime })\equiv \frac{1}{\sqrt{2g_0}}\Vert P_{0}-P_{0}^{\prime }\Vert _{2},
\end{equation}%
whence, keeping only the lowest non-vanishing order, we have 
\begin{eqnarray}
d^{2}(P_{0}(\mathbf{x}),P_{0}(\mathbf{x}+\mathrm{d}\mathbf{x})) &=&\frac{1}{2g_0}\Vert
P_{0}(\mathbf{x}+\mathrm{d}\mathbf{x})-P_{0}(\mathbf{x})\Vert _{2}^{2} 
\notag \\
&=&\frac{1}{2g_0}\mathrm{Tr}[(\mathrm{d}P_{0}(\mathbf{x})+\frac{1}{2}\mathrm{d}^{2}P_{0}(%
\mathbf{x}))^{2}]  \notag \\
&=&\frac{1}{2g_0}\mathrm{Tr}[\mathrm{d}P_{0}(\mathbf{x})\mathrm{d}P_{0}(\mathbf{x})]\notag\\
&=&\frac{1}{2g_0}\mathrm{Tr}[\partial _{i}P_{0}\mathrm{d}{\mathbf{x}}^{i}\partial _{j}P_{0}%
\mathrm{d}{\mathbf{x}}^{j}]  \notag \\
&=&\mathbf{g}_{ij}\mathrm{d}{\mathbf{x}}^{i}\mathrm{d}{\mathbf{x}}^{j},
\label{d^2}
\end{eqnarray}%
with the metric tensor as defined in Eq.~(\ref{g2}). \emph{Thus the
adiabatic metric tensor is precisely the metric over the Grassmannian
manifold defined by the ground state projectors}.

%%%%%%%%%%%%%%%%%%%%%%%%%%%%%%%%%%%%%%%%%%%%%%%%%%%%%%%%%%%
\subsubsection{Adiabatic parallel transport}

In this subsection we wish to define a notion of adiabatic parallel
transport. We start with the standard purification \cite{Uhlmann1,Uhlmann2} 
\begin{equation}
W=P_{0}U  \label{W}
\end{equation}%
of $P_{0}$, where $U$ is an arbitrary unitary acting on $\mathcal{H}$, so
that $P_{0}=WW^{\dag }$. Here $W$ is considered a vector in a larger
(extended) Hilbert space $\mathcal{H}_{\mathrm{ext}}$, i.e., a pure state
whose reduction yields (the density matrix) $P_{0}$. The Hilbert space $%
\mathcal{H}_{\mathrm{ext}}$ is equipped with the the Hilbert-Schmidt inner
product%
\begin{equation}
\langle A,B\rangle :=\mathrm{Tr}[A^{\dag }B].
\label{HS-inprod}
\end{equation}%
Given $P_{0}$, the fiber of all purifications sitting on the unit sphere $%
\mathcal{S}(\mathcal{H}_{\mathrm{ext}}):=\{W\in \mathcal{H}_{\mathrm{ext}}:$ 
$\langle W,W\rangle =1\}$ of $\mathcal{H}_{\mathrm{ext}}$ is the Stiefel
manifold of orthonormal $g_{0}$-frames of $\mathcal{H}_{\mathrm{ext}}$,
where $\mathrm{Tr}[P_{0}]=g_{0}$ (i.e., the set of ordered $g_{0}$-tuples of
orthonormal vectors in $\mathcal{H}_{\mathrm{ext}}$). The gauge
transformation (\ref{W}) means that the fiber admits the unitaries of $%
\mathcal{H}$ as right multipliers. Informally, the $U$s act as arbitrary
\textquotedblleft phases\textquotedblright\ associated with $P_{0}$.

Starting with a curve of (unnormalized) density operators $s\mapsto P_{0}(s)$ and one of
its purifications%
\begin{equation}
s\mapsto W(s)\quad P_{0}(s)=W(s)W^{\dag }(s),
\end{equation}%
the length $\ell _{U}[s]=\int_{0}^{s}\langle \dot{W}(s^{\prime }),\dot{W}%
(s^{\prime })\rangle \mathrm{d}s^{\prime }$ of the curve in $\mathcal{H}_{%
\mathrm{ext}}$ is not invariant against gauge transformations (\ref{W}). The
Euler equations for the variational problem $\ell \lbrack s]:=\inf_{U}\ell
_{U}[s]$, i.e., for the geodesic are \cite{Uhlmann1,Uhlmann2}%
\begin{equation}
W^{\dag }\mathrm{d}W=\mathrm{d}W^{\dag }W,
\end{equation}%
also known as the Uhlmann parallel transport\ condition. Substituting $%
W^{\dag }=U^{\dag }P_{0}$ and $\mathrm{d}W=(\mathrm{d}P_{0})U+P_{0}\mathrm{d}%
U$ yields the condition%
\begin{equation}
U^{\dag }P_{0}((\mathrm{d}P_{0})U+P_{0}\mathrm{d}U)=(U^{\dag }\mathrm{d}%
P_{0}+(\mathrm{d}U^{\dag })P_{0})P_{0}U,
\end{equation}%
which, using $U\mathrm{d}U^{\dag }=-(\mathrm{d}U)U^{\dag }$, reduces to 
\begin{equation}
P_{0}(\mathrm{d}U)U^{\dag }+(\mathrm{d}U)U^{\dag }P_{0}=[\mathrm{d}%
P_{0},P_{0}]  \label{fiber1}
\end{equation}%
on the vector bundle over the Grassmannian $\mathcal{G}_{N,g_{0}}$. Here $%
U=U(s)$ is a general unitary undergoing parallel transport as $s\mapsto
P_{0}(s)$. We now seek those unitaries $U$ which in addition to parallel
transport, also satisfy adiabaticity.

To this end let $J(s)$ be the infinitesimal generator of $U(s)$, i.e.,%
\begin{equation}
i\partial _{s}U(s)=TJ(s)U(s).
\end{equation}%
Substituting this expression into Eq.~(\ref{fiber1}) we obtain 
\begin{eqnarray}
P_{0}J+JP_{0} &=&i[\partial _{s}P_{0},P_{0}]/T  \label{PJ+JP1} \\
&=&H_{\text{ad}}-H,  \label{PJ+JP2}
\end{eqnarray}%
where in the second line we used Eq.~(\ref{Had-H}). Thus, $U$ satisfies 
\emph{adiabatic parallel transport} if in addition to being a solution to
the parallel transport condition (\ref{fiber1}) its generator also satisfies
the adiabaticity condition 
\begin{equation}
P_{0}J+JP_{0}=0.  \label{ad-trans-def}
\end{equation}

What is the generator $J$ which satisfies this last condition? Using Eqs.~(%
\ref{PdotP}) and (\ref{Eb}) for the nondegenerate case we 
obtain
\begin{eqnarray}
-iT(P_{0}J+JP_{0}) &=&[\dot{P}_{0},P_{0}]\nonumber\\ & =& -\frac{1}{H-E_{0}}\dot{H}%
P_{0}+P_{0}\dot{H}\frac{1}{H-E_{0}}  \notag \\
&=&\sum_{n>0}\frac{P_{0}\dot{H}|\Phi _{n}\rangle \langle
\Phi _{n}|-|\Phi _{n}\rangle \langle \Phi _{n}|\dot{H}P_{0}}{E_{n}-E_{0}}.\nonumber\\  \label{PJ+JP}
\end{eqnarray}%
Taking matrix elements we find $\langle \Phi _{0}|J|\Phi _{0}\rangle =0$ and 
$-iT\langle \Phi _{0}|J|\Phi _{k}\rangle =\frac{1}{E_{k}-E_{0}}\langle \Phi
_{0}|\dot{H}|\Phi _{k}\rangle ,$ while the matrix elements of $J$ between
the excited states are unspecified, so that 
\begin{equation}
J=\frac{i}{T}\sum_{n>0}\dfrac{\langle \Phi _{0}|\partial _{s}H|\Phi
_{n}\rangle }{E_{n}-E_{0}}|\Phi _{0}\rangle \langle \Phi _{n}|+\text{H.c.}%
+J_{\bot },  \label{J}
\end{equation}%
where $J_{\bot }$ is an arbitrary operator satisfying $J_{\bot
}=Q_{0}J_{\bot }Q_{0}$. 

Instead of trying to obtain perfect adiabaticity ($H_{\text{ad}}=H$) we can
settle for an approximation. Noting that Eqs.~(\ref{eps2}) and (\ref{PJ+JP1}) imply 
\begin{eqnarray}
\epsilon (s) &=& T\int_{0}^{s}\Vert P_{0}(s^{\prime })J(s^{\prime })+J(s^{\prime
})P_{0}(s^{\prime })\Vert _{2}\mathrm{d}s^{\prime },\nonumber\\
& =& \int_0^{s}\sqrt{2g_0\mathbf{g}_{ij}(\mathbf{x})\dot{\mathbf{x}}^i \dot{\mathbf{x}}^j}\mathrm{d}s'
\end{eqnarray}%
it follows that minimizing $\epsilon $, 
or equivalently finding the adiabatic geodesic, 
endows the ``phase'' $U$ of $P_{0}$ with an adiabatic characteristic
which is compatible with the Uhlmann parallel transport condition. 
Thus, we have shown that
the metric tensor $\mathbf{g}$ emerges naturally
also from the notion of adiabatic parallel transport.

%%%%%%%%%%%%%%%%%%%%%%%%%%%%%%%%%%%%%%%%%%%%%%%%%%%%%%%%%%%
\subsubsection{Bures metric}

There is also a straightforward connection between our metric and the
Bures metric \cite{Hayashi:book}. For two arbitrary density matrices
$\rho_1$ and $\rho_2$, the Bures distance is defined as
\begin{eqnarray}
d^2_{\text{Bures}}(\rho_1,\rho_2) \equiv 1-F(\rho_1,\rho_2),
\end{eqnarray}
where $F(\rho_1,\rho_2)\equiv
\mathrm{Tr}[(\rho_{1}^{1/2}\rho_2\rho_1^{1/2})^{1/2}]$ is the fidelity
between these two states \cite{Nielsen-book,bures-comment}. When the density
matrices depend on a parameter $\mathbf{x}$, the infinitesimal
distance
$d^2_{\text{Bures}}\left(\rho(\mathbf{x}),\rho(\mathbf{x}+\mathrm{d}\mathbf{x})
\right)$ can be shown to be \cite{Information2}
\begin{eqnarray}
d^2_{\text{Bures}}\left(\rho(\mathbf{x}),\rho(\mathbf{x}+\mathrm{d}\mathbf{x}) \right) = \mathrm{Tr}[\rho(\mathbf{x})\mathcal{L}^2(\mathbf{x})],
\label{b-metric}
\end{eqnarray} 
where $\mathcal{L}(\mathbf{x})$ is the ``symmetric logarithmic
derivative,'' (SLD) defined via
\begin{eqnarray}
\mathrm{d}\rho(\mathbf{x}) = \frac{1}{2}\bigl(\mathcal{L}(\mathbf{x})\rho(\mathbf{x}) +\rho(\mathbf{x})\mathcal{L}(\mathbf{x}) \bigr).
\label{SLD}
\end{eqnarray}

From the property $P_0^2=P_0$ we obtain
\begin{eqnarray}
\mathrm{d}P_0(\mathbf{x})=\mathrm{d}P_0(\mathbf{x}) P_0(\mathbf{x})+ P_0(\mathbf{x}) \mathrm{d}P_0(\mathbf{x}).
\end{eqnarray}
and hence [see Eq.~(\ref{PdPP})]
\begin{eqnarray}
\mathrm{d}g_0=\mathrm{Tr}[\mathrm{d}P_0]= 2\mathrm{Tr}[P_0 \mathrm{d}P_0]= 2\mathrm{Tr}[P_0
  \mathrm{d}P_0 P_0]=0,
\end{eqnarray}
i.e., the degeneracy is constant. Thus if $\rho(\mathbf{x})\equiv
  P_0(\mathbf{x})/g_0$, then $\mathrm{d} 
  [P_0(\mathbf{x})/g_0] = P_0(\mathbf{x})/g_0 \mathrm{d} P_0 +
  \mathrm{d} P_0 P_0(\mathbf{x})/g_0$, and the
definition of the SLD [Eq.~(\ref{SLD})] yields
\begin{eqnarray}
\mathcal{L}(\mathbf{x})= 2\mathrm{d}P_0(\mathbf{x}).
\end{eqnarray}
Inserting this back into Eq.~(\ref{b-metric}) results in
\begin{eqnarray}
d^2_{\text{Bures}}\left(P_0(\mathbf{x}),P_0(\mathbf{x}+\mathrm{d}\mathbf{x}) \right) &=&\frac{4}{g_0} \mathrm{Tr}[P_0(\mathbf{x})(\mathrm{d}P_0(\mathbf{x}))^2]\nonumber\\
& \equiv &  \mathbf{g}_{ij}^{\text{Bures}}(\mathbf{x}) \mathrm{d}\mathbf{x}^i \mathrm{d}\mathbf{x}^j,
\end{eqnarray}
where
\begin{eqnarray}
 \mathbf{g}_{ij}^{\text{Bures}}(\mathbf{x})=\frac{4}{g_0} \mathrm{Tr}[\partial_i P_0(\mathbf{x}) \partial_j P_0(\mathbf{x})].
\end{eqnarray}
By comparison with Eq.~(\ref{g2}), we obtain
\begin{eqnarray}
\mathbf{g}_{ij}^{\text{Bures}} = 8 \mathbf{g}_{ij}.
\end{eqnarray}
We note
that the Bures metric is also connected to ``quantum Fisher
information tensor,'' which plays a principal role in quantum
estimation theory
\cite{Information2,Hayashi:book,Paris,estimation}. In fact the
Bures metric is (up to an unimportant constant multiplicative factor)
equal to the Fisher tensor. Therefore, the adiabatic metric is the
quantum Fisher metric. The role of the metric $\mathbf{g}$ in quantum
estimation theory is thus highlighted naturally this way. 

%%%%%%%%%%%%%%%%%%%%%%%%%%%%%%%%%%%%%%%%%%%%%%%%%%%%%%%%%%%
\subsection{Comparison of adiabatic metrics}
\label{Comp}

In adiabatic evolution (as well as in adiabatic quantum computation)
$\delta$ and $T$ are the
primary objects of interest. Our method for obtaining the
metric $\mathbf{g}$ here is based on minimizing an upper bound on
the adiabatic
error $\delta$ for a given evolution time $T$. In Ref.~\cite{QAB} we pursued a
complementary route and proposed a different metric, 
\begin{eqnarray}
 \widetilde{\mathbf{g}}_{ij}(\mathbf{x})=\mathrm{Tr}[\partial_i H(\mathbf{x}) \partial_j H(\mathbf{x})]/\Delta^4(\mathbf{x}),
 \label{gtilde}
\end{eqnarray}
derived from minimizing a time functional inspired by the traditional
 adiabatic condition. We called this the ``quantum adiabatic brachistochrone''.

The major difference between these two metrics is in their distinct gap dependence. This can be
understood, for example, by noting that
\begin{eqnarray}
|\mathbf{g}_{ij}|\leq \frac{\Vert \partial_i H \partial_jH\Vert_{1}}{\min_{s}\Delta^2},
\label{g-bnd1}
\end{eqnarray}
whereas
\begin{eqnarray}
|\widetilde{\mathbf{g}}_{ij}|\leq \frac{\Vert \partial_i H \partial_j H\Vert_{1}}{\min_{s}\Delta^4},
\label{g-bnd2}
\end{eqnarray}
where $\Vert X \Vert_1\equiv \mathrm{Tr}[\sqrt{X^{\dag}X}]=\sum_{i}\sigma_i(X)$ is the trace 
norm \cite{Bhatia:book} (see Appendix~\ref{app:g-bounds} for the proof). Thus, the metric $\mathbf{g}$ has a quadratically less dependence on the inverse gap. This may imply different behaviors for these metrics and their corresponding curvatures; hence they are essentially distinct.

%%%%%%%%%%%%%%%%%%%%%%%%%%%%%%%%%%%%%%%%%%%%%%%%%%%%%%%%%%%
\subsection{Strategies for reducing the adiabatic error and their effect on geometry}
\label{Red}

Considering that $\mathbf{g}$ is related to minimizing the upper bound on $\delta$, it is useful to briefly recall how $\delta$ scales with $T$ and how this scaling may be improved. 

Rigorous proofs of the adiabatic theorem---based on successive integration by parts of $%
\Omega$---state that if $\{H(s)\}$ is a family of $C^k$ ($k$ times
continuously differentiable) interpolations/paths with bounded $\Vert
\partial^l_s{H}\Vert$ ($l\in\{1,\ldots,k\}$) and compactly supported $%
\partial_s H$ over $s\in(0,1)$, then $\delta=\mathcal{O}(1/T^{2(k-1)})$ \cite{Avron:871,Avron:872,JRS}. If these assumptions are supplemented with that
of analyticity of $H(s\in\mathbb{C})$ in a small strip around the real
time axis, and if in addition
\begin{eqnarray}
\partial^l_s H(0)=\partial^l_s H(1)=0~~\forall l\leq k,
\end{eqnarray}
the result is an
\textit{exponentially} smaller error 
\begin{eqnarray}
\delta=\mathcal{O}( e^{-cT}),
\end{eqnarray}
where
$c\equiv \min_s\Delta^3/\max_s\Vert \partial_s H\Vert^2$ (up to an
$\mathcal{O}(1)$ prefactor) \cite{LRH}.

Our path---as the solution to the second-order differential equation
(\ref{EL-eq})---minimizes $\epsilon$ rather than $\delta$, which is
not necessarily compatible with the boundary conditions $\partial^l_s
H(\{0,1\})=0$. Thus, in principle, there remains room for further
optimization of the path for $\delta$ beyond what is captured by simply
minimizing its upper bound $\epsilon(s)\mathcal{O}(1/T)$ \cite{Avron:871,Avron:872}. Such finer
optimizations, however, may not always result in a Riemannian geometry because the corresponding functionals and Euler-Lagrange equations would depend on higher derivatives of $H$.

\section{Connection to quantum phase transitions}
\label{QPT}

The other physically-important aspect of our geometric formulation
emerges from the observation that the metric $\mathbf{g}$ also
arises naturally as the underlying geometry of QPTs. QPTs take place
at zero temperature \cite{Sachdev-book}, where the system is in
principle in its ground state. Such phase transitions exhibit peculiar
behaviors and ``orders,'' radically different from their thermal
counterparts. In particular, in contrast to thermal phase transitions,
the standard paradigm of the Landau-Ginzburg symmetry-breaking
mechanism \cite{Goldenfeld-book,Sachdev-book} fails to explain the
underlying physics of some QPTs. In fact, defining an appropriate
\textit{local} ``order parameter''---an essential ingredient of the
Landau-Ginzburg theory---is not straightforward for a quantum critical system; some QPTs, such as those involving ``topological order,'' provably do not admit any local order parameter
\cite{Top-Order1,Top-Order2}. Additionally, tracking singularities of
the ground-state energy cannot always foreshadow QPTs; quantum systems
with matrix-product states may elude this test \cite{Cirac-MPS}.  

Notwithstanding the above subtleties with identifying QPTs, it has
recently been shown that the simple notion of the ``ground-state
fidelity'' is remarkably successful in signaling QPTs
\cite{Zanardi-Paunkovic:06,Zanardi-prl:07}. This can be understood by
noting that since QPTs take place at zero temperature, in which the
system is in its ground state, quantum criticality should be
identifiable by ground state properties. Specifically, the ground
states right before and right after a quantum critical point are
expected to have very little overlap. In this manner ground-state
fidelity may be considered as a natural, fairly general order
parameter for quantum critical systems, irrespectively of their
internal symmetries \cite{Zanardi-prl:07}. We shall discuss this
feature in more detail below.

\subsection{Metric tensor for QPTs}
\label{QPT-metricg}

Here we derive the metric attributed to QPTs for the case of degenerate ground states as a natural extension of the similar metric proposed for the nondegenerate case \cite{Zanardi-prl:07}.

In the degenerate case we should work with the ground-state projector $P_0(\mathbf{x})$. A variation in the properties of $P_{0}(\mathbf{x})$, caused by the change $\mathbf{x}\rightarrow \mathbf{x}+\mathrm{d}\mathbf{x}$ in the Hamiltonian parameters, can be captured by the order parameter chosen to be the
operator fidelity of $P_{0}(\mathbf{x})$ and
$P_{0}(\mathbf{x}+\mathrm{d}\mathbf{x})$ relative to, e.g.,
$\varrho=I_{g_0}/g_0$ ($I_{g_0}$ is the $g_0\times g_0$ identity matrix),
\begin{eqnarray}
\hskip-2mm f_{\varrho}\bigl(P_0(\mathbf{x}),P_0(\mathbf{x}+\mathrm{d}\mathbf{x})\bigl) &=& \langle P_{0}(\mathbf{x}),P_{0}(\mathbf{x}+\mathrm{d}\mathbf{x})\rangle_{\varrho} \nonumber\\
& =& 1- \mathbf{G}_{ij}(\mathbf{x})\mathrm{d}\mathbf{x}^i\mathrm{d}\mathbf{x}^j, 
\label{geometric tensor}
\end{eqnarray}
in which the Hermitian matrix
\begin{eqnarray}
\mathbf{G}_{ij}\equiv\frac{1}{g_0}\mathrm{Tr}[P_0 (\partial_i P_0) (\partial_j P_0)P_0]
\label{g_tensor}
\end{eqnarray}
is the ``geometric tensor" for the degenerate case (see Appendix~\ref{app:G-proof} for the proof). Thus the information about the criticality of the quantum system is contained in the $\mathbf{G}$ tensor. Note that in the nondegenerate case ($g_0=1$) $\mathbf{G}_{ij}$ reduces to
\begin{eqnarray}
\mathbf{G}_{ij} & = & 
\langle \partial_i\Phi_0|\partial_j\Phi_j\rangle - \langle \partial_i \Phi_i|\Phi_0\rangle \langle \Phi_0|\partial_j\Phi_0\rangle.
\end{eqnarray}

Accordingly, a Riemannian QPT metric tensor can be defined through 
\begin{eqnarray}
\mathbf{g}^{\text{QPT}}_{ij}(\mathbf{x}) &\equiv& \mathrm{Re}[\mathbf{G}_{ij}(\mathbf{x})]=\frac{1}{2g_0}\mathrm{Tr}[\partial_i P_0 \partial_j P_0] \nonumber\\
& = & \mathbf{g}_{ij},
\label{Gg}
\end{eqnarray} 
where we used the same trick as that used in arriving at
Eq.~(\ref{temp-1}). Therefore, the
QPT metric tensor $\mathbf{g}^{\text{QPT}}$ is the same as the adiabatic quantum evolution
metric $\mathbf{g}$ defined in Eq.~(\ref{g2}). 

\subsection{Quantum critical scaling of the QPT metric tensor}
\label{g-criticality}

The critical behavior of a quantum system with a degenerate ground
state can be characterized by the metric tensor $\mathbf{g}$. This is
already evident from the fact that the divergence of
$\mathbf{g}^{\text{QPT}}$ is a \textit{sufficient} condition for
signaling a quantum criticality. To further elaborate on this
connection, we follow Ref.~\cite{VZ} and obtain the scaling of the
geometric tensor (\ref{g_tensor})
\begin{eqnarray}
 \mathbf{G}_{ij}&=&\frac{1}{g_0}\mathrm{Tr}\left[ P_{0}(\partial _{i}H)\left( \frac{1}{H-E_{0}}\right)
^{2}(\partial _{j}H)P_{0}\right] \label{eq-**} \\
 &=& \frac{1}{g_0} \sum_{n>0}\sum_{\alpha,\eta=1}^{g_0,g_n}
 \frac{\langle \Phi_0^{\alpha}|\partial_i H|\Phi_n^{\eta} \rangle
   \langle \Phi_n^{\eta}|\partial_j
   H|\Phi_0^{\alpha}\rangle}{(E_n-E_0)^2},
\label{eq-*}
\end{eqnarray}
and via Eq.~(\ref{Gg}) also for $\mathbf{g}_{ij}$. 
For simplicity we restrict ourselves only to gapped quantum systems
with second-order QPTs. Thus, in a critical region $\mathbf{x}\approx
\mathbf{x}_c$, the correlation length $\xi$  and the gap $\Delta$
exhibit the following scalings 
\begin{eqnarray}
 & \xi\sim|\mathbf{x}-\mathbf{x}_c|^{-\nu},~~\Delta\sim|\mathbf{x}-\mathbf{x}_c|^{z\nu},
 \label{xi-scaling}
\end{eqnarray}
with the critical exponents $\nu>0$ and $z\nu$, where $z>0$ is the dynamical
exponent \cite{Sachdev-book}.
The geometric tensor $\mathbf{G}$
has an integral representation
which not only facilitates the derivation of the scaling relation for
$\mathbf{G}$, but also enables an interpretation for $\mathbf{G}$ in
terms of correlation (or response) functions.
Indeed, as shown in Appendix~\ref{app:g-integral}, Eq.~(\ref{eq-*}) can be expressed as 
\begin{eqnarray}
\mathbf{G}_{ij}&=&\frac{1}{g_0}\int_{0}^{\infty} \mathrm{d}\tau \, \tau e^{-p\tau}\Big(\mathrm{Tr}[P_0\partial_i H_{\tau} \partial_j H]\nonumber \\
 &&- \frac{1}{g_{0}}\mathrm{Tr}[P_0\partial_i H]\mathrm{Tr}[P_0\partial_j H] \Big)\Big |_{p=0},
 \label{G-integral}
\end{eqnarray}
with $\partial_i H_{\tau} \equiv e^{\tau H}\partial_i H e^{-\tau H}$. 

Now we make some generic assumptions about the Hamiltonian $H$. First, let $\partial_i H$ be a \textit{local} operator; that is, one can write
\begin{eqnarray}
 \partial_{i}H = \sum_{\mathrm{y}} h_i(\mathrm{y}),
\end{eqnarray}
in which $\mathrm{y}$ labels the spatial region over which the local
operator $h_i(\mathrm{y})$ has support. Second, the $h_i(\mathrm{y})$
operators have well-defined scaling dimensions
$\alpha_i$ near the quantum critical point
$\mathbf{x}_c$, such that if 
\begin{eqnarray}
 & \mathrm{y}\rightarrow a \mathrm{y},~~\tau\rightarrow a^{z}\tau,
\end{eqnarray}
 for $a>0$, we obtain 
\begin{eqnarray}
  h_i(\mathrm{y}) \rightarrow a^{-\alpha_i}h_i(\mathrm{y}).
\label{crit-dim}
\end{eqnarray}
Under these transformations, Eq.~(\ref{G-integral}) yields the following scaling for the rescaled geometric tensor in the thermodynamic limit 
\begin{eqnarray}
 \frac{1}{L^d} \mathbf{G}_{ij}\rightarrow a^{-\kappa_{ij}} \frac{1}{L^d}\mathbf{G}_{ij},
\end{eqnarray}
where 
\begin{eqnarray}
 \kappa_{ij}\equiv \alpha_i+\alpha_j-2z-d.
\label{eq:kappa}
\end{eqnarray}
Here, $L$ is the linear size of the system and $d$ is its spatial
dimension. From Eq.~(\ref{xi-scaling}), we obtain
$|\mathbf{x}-\mathbf{x}_c|\sim \xi^{-1/\nu}$; i.e., the scaling
dimension of the Hamiltonian parameter $\mathbf{x}$ is
$1/\nu$. Following standard scaling analysis arguments, the scaling behavior of the metric tensor (recall that $\mathbf{g}=\mathrm{Re}[\mathbf{G}]$) in the off-critical limit $\xi\ll L$ is  
\begin{eqnarray}
 \mathbf{g}_{ij}(\mathbf{x}\approx \mathbf{x}_c) \approx L^{d} |\mathbf{x}-%
\mathbf{x}_c|^{\nu\kappa_{ij}}. 
 \label{g-scaling}
\end{eqnarray}
Moreover, in the critical region, where
$\xi\gg L\gg $ the spacing between adjacent particles on the system
lattice, in addition to the regular extensive scaling $L^d$, the
finite-size scaling of the metric is $\mathbf{g}_{ij}\sim L^{d-\kappa_{ij}}$%
, which could be extensive, subextensive, or superextensive [$\kappa=0$,
positive, or negative, respectively]. We also remark that there exist
models, exhibiting quantum topological order, in which the
critical $\mathbf{g}$ scales logarithmically, e.g., $\mathbf{g}\sim\ln|\mathbf{x}-\mathbf{x}_c|$ 
\cite{logg1,logg2,logg3}.
 
%%%%%%%%%%%%%%%%%%%%%%%%%%%%%%%%%%%%%%%%%%%%%%%%%%%%%

\section{Adiabatic geodesics}
\label{secIII}

In this section
we solve the geodesic equation (\ref{EL-eq})
analytically for some specific examples. Note
that since the eigenprojections do not depend on $\mathrm{Tr}[H]$,
Eq.~(\ref{EL-eq}) corresponds to an underdetermined system of coupled
second-order differential equations. 
This can be seen more clearly by adopting a new parametrization (i.e.,
coordinate system) $\mathbf{y}(\mathbf{x})$ for the Hamiltonian such that $H%
\bigl(\mathbf{y}(\mathbf{x})\bigr)= \mathrm{y}^1(\mathbf{x})\openone +
H^{\prime }\bigl( \mathrm{y}^2(\mathbf{x}),\ldots, \mathrm{y}^M(\mathbf{x}) %
\bigr)$, in which $\mathrm{y}^1=\mathrm{Tr} [H]/N$ and $H^{\prime }=H-%
\mathrm{Tr}[H]\openone/N$. Since $P_0(\mathbf{y})$ does not depend on
$\mathrm{y}^1$, the metric $\mathbf{g}(\mathbf{y})$ does not depend on
this parameter either. Independence from $\mathrm{y}^1$ translates in
terms of $\mathbf{x}$ into the statement that only $M-1$ equations in the system 
(\ref{EL-eq}) are independent. 

%%%%%%%%%%%%%%%%%%%%%%%%%%%%%%%%%%%%%%%%%%%%%%%%%%%%%%%%%%%
\subsection{Examples}
\label{examples}

\subsubsection{Deutsch-Jozsa algorithm}

In the Deutsch-Jozsa algorithm \cite{Deutsch:92} one is given an oracle that calculates
a function $f:\{0,1\}^n\mapsto \{0,1\}$. The promise is that $f$ is
either ``constant'' or ``balanced,'' meaning respectively that,
$f(\mathbf{i})=f(\mathbf{i}^{\prime })~\forall
\mathbf{i},\mathbf{i}^{\prime }$ or $f(\text{half of all
}\mathbf{i}\text{'s})=0$ \cite{Nielsen-book}. The objective is to
conclude whether $f$ is constant or balanced. The Deutsch-Jozsa
algorithm finds the answer by querying the oracle only once,
while classical deterministic algorithms require a number of queries that is exponential in $n$.

An adiabatic version of this algorithm was
introduced in Ref.~\cite{Lidar-DJ}. We consider the \emph{unitary interpolation} Hamiltonian \cite{Siu:05}
\begin{eqnarray}
 H\bigl( \mathrm{x}(s)\bigr) = \widetilde{V}( \mathrm{x}(s)\bigr) H_0 
\widetilde{V}^{\dag}( \mathrm{x}(s)\bigr),  \label{DJ}
\end{eqnarray}
where $\widetilde{V}\bigl( \mathrm{x}(s)\bigr)=e^{i\frac{\pi}{2} \mathrm{x}%
(s)G}$, in which the Hermitian/unitary $G$ operator is defined by
$G|\mathbf{i}\rangle=(-1)^{f(\mathbf{i})}|\mathbf{i}\rangle$. Here
$H_0$ is chosen such that $|\Phi_0(0)\rangle=|+\rangle^{\otimes n}=
2^{-n/2}\sum_{\mathbf{i}=0}^{2^n-1}| \mathbf{i}\rangle$ is its ground
state, e.g.,
\begin{eqnarray}
& H_0=h_0 \sum_{k=1}^n|-\rangle_k\langle -|,
\end{eqnarray}
where $|\pm\rangle=(|0\rangle\pm|1\rangle)/\sqrt{2}$, $\sigma_z=|0\rangle\langle0|-|1\rangle\langle 1|$ is a
Pauli matrix, and $h_0>0$ is an energy scale. The boundary conditions
are chosen
as $(\mathrm{x}_0,\mathrm{x}_1)=(0,1)$ such that $H(0)=H_0$ and
$H(1)=GH_0G^{\dag}$; the latter guarantees that $|\Phi_0(1)\rangle
=G|\Phi_0(0)\rangle$ is the ground state of $H(1)$.

From Eq.~(\ref{DJ}) it is seen that $|\Phi_0(s)\rangle = \widetilde{V}(s)|\Phi_0(0)\rangle$, whence we obtain 
\begin{eqnarray}
&\hskip-3mm P_0\bigl(\mathrm{x}(s)\bigr) =
  2^{-n}\sum_{\mathbf{i},\mathbf{i}'=0}^{2^n-1}e^{i\frac{\pi}{2}\mathrm{x}(s)[(-1)^{f(\mathbf{i})}-(-1)^{f(\mathbf{i}')}]}|\mathbf{i}\rangle \langle \mathbf{i}'|, 
\notag 
\\
& \partial_{\mathrm{x}}P_0 \bigl(\mathrm{x}(s)\bigr) = i\frac{\pi}{2}[G,P_0\bigl(\mathrm{x}(s)\bigr)].
\end{eqnarray}
A straightforward calculation then yields 
\begin{eqnarray}
 \mathbf{g}& =& \mathrm{Tr}[\bigl(\partial_{\mathrm{x}}P_0(\mathrm{x}) \bigr)^2]\nonumber\\
& = & \frac{\pi^2}{2}\left(\mathrm{Tr}[P_0(\mathrm{x})G^2]-\mathrm{Tr}[\bigl( P_0(\mathrm{x})G\bigr)^2]\right)\nonumber
\\
& =& \frac{\pi^2}{2}\left[1-2^{-2n}\bigl( \sum_{\mathbf{i}=0}^{2^n-1}e^{i\pi f(\mathbf{i})}\bigr)^2\right].
\end{eqnarray}
Since $\mathbf{g}$ is independent of $\mathrm{x}(s)$, the geodesic equation (\ref{EL-eq}) reduces to $\ddot{\mathrm{x}}=0$, whence the geodesic is simply
\begin{eqnarray}
 \mathrm{x}(s)=s,
\end{eqnarray}
which corresponds to a rotation of the initial Hamiltonian $H_0$ at a
constant rate.

\subsubsection{Projective Hamiltonians}

Consider the following Hamiltonian:
\begin{eqnarray}
& H\bigl(\mathbf{x}(s)\bigr)=\mathrm{x}^1(s)P^{\perp}_{\mathbf{a}} + \mathrm{%
x}^2(s) P^{\perp}_{\mathbf{b}},
\label{bi-proj}
\end{eqnarray}
where $P^{\perp}_{\mathbf{a}}=\openone-|\mathbf{a}\rangle \langle \mathbf{a}%
| $, for a given $|\mathbf{a}\rangle\in\mathcal{H}$ (similarly for $%
P^{\perp}_{\mathbf{b}}$), $\langle\mathbf{a}|\mathbf{b}\rangle$ is a given
function of $N$, and the boundary conditions are $\mathbf{x}_0=(1,0)$ and $%
\mathbf{x}_1=(0,1)$. This Hamiltonian may represent the adiabatic
preparation of an unknown (``hard") state $|\mathbf{b}\rangle$ from the
supposedly known (``simple") initialization $|\mathbf{a}\rangle$, provided
that one has access to the ``oracle" $P^{\perp}_{\mathbf{b}}$ \cite{QAB}. An
important instance of this class is Grover's Hamiltonian for search of a
``marked'' item among $N$ unsorted items \cite{Grover:97a}
(generalized to arbitrary initial amplitude distributions in
Refs.~\cite{Biham:99,Biham:00}), where
$|\mathbf{a}\rangle=\sum_{k=0}^{N-1}|k\rangle/\sqrt{N}$ and
$|\mathbf{b}\rangle=|m\rangle$, for $m\in\{0,\ldots,N-1\}$. A
successful adiabatic version of this algorithm was first described in
Ref.~\cite{RolandCerf}.

Since the Hamiltonian (\ref{bi-proj}) is effectively two-dimensional
over the span of the vectors $|\mathbf{a}\rangle$ and $|\mathbf{b}\rangle$, it
can be diagonalized analytically. 
Indeed, given $|\mathbf{a}\rangle$, we have the
freedom to choose $N-1$ vectors $\{|\mathbf{a}^{\perp}_i\rangle\}_{i=1}^{N-1}$
at will such that together with $|\mathbf{a}\rangle$ they constitute an orthonormal basis for
$\mathcal{H}$. I.e., $\langle\mathbf{a}|\mathbf{a}^{\perp}_i\rangle=0$
and
$\langle\mathbf{a}^{\perp}_i|\mathbf{a}^{\perp}_j\rangle=\delta_{ij}$.
Thus we can decompose $|\mathbf{b}\rangle= \alpha_0|\mathbf{a}\rangle+
\sum_{i=1}^{N-1}\alpha_i |\mathbf{a}^{\perp}_i\rangle$. Utilizing the
freedom in choosing $\{|\mathbf{a}^{\perp}_i\rangle\}$ (up to the
orthonormality condition), we can always rotate them such that 
$\alpha_{i>1}=0$. In this case, we have 
\begin{eqnarray}
& |\mathbf{b}\rangle= \alpha_0|\mathbf{a}\rangle + \alpha_1 |\mathbf{a}^{\perp}_1\rangle,
\label{b-exp}
\end{eqnarray}
where $\alpha_0=\langle \mathbf{a}|\mathbf{b}\rangle$ and
$\alpha_1=\langle \mathbf{a}^{\perp}_1| \mathbf{b}\rangle$ (or more
explicitly: $\alpha_1 = e^{i\phi_1}\sqrt{1-|\langle
  \mathbf{a}|\mathbf{b}\rangle|^2}$, for some arbitrary
$\phi_1\in[0,2\pi)$). 

Expanding Eq.~(\ref{bi-proj}) in the $\{|\mathbf{a}\rangle,
|\mathbf{a}^{\perp}_i\rangle \}_{i=1}^{N-1}$ basis and using
Eq.~(\ref{b-exp}) yields
\begin{eqnarray}
 H(\mathbf{x}) &=& \left(\begin{array}{cc}
   \mathrm{x}^2(1-|\alpha_0|^2)& -\mathrm{x}^2 \alpha_0\bar{\alpha}_1
   \\ -\mathrm{x}^2 \bar{\alpha}_0 \alpha_1 & \mathrm{x}^1
   +\mathrm{x}^2|\alpha_0|^2 \end{array}\right)\nonumber\\ 
&& \oplus (\mathrm{x}^1+\mathrm{x}^2) I_{\{2,\ldots,N-1\}},
\label{H-matrixform}
\end{eqnarray}
where we have used the completeness of the basis to write $I =
|\mathbf{a} \rangle\langle\mathbf{a}|+\sum_{i=1}^{N-1}
|\mathbf{a}^{\perp}_i \rangle \langle\mathbf{a}^{\perp}_i|$,
$I_{\{2,\ldots,N-1\}} \equiv \sum_{i=2}^{N-1} |\mathbf{a}^{\perp}_i
\rangle \langle\mathbf{a}^{\perp}_i|$, and the matrix on the right hand side is written in the
$\{|\mathbf{a}\rangle,|\mathbf{a}^{\perp}_1\rangle\}$ (sub-)basis. It then follows
from Eq.~(\ref{H-matrixform}) that
the spectrum of $H$ consists of the two nondegenerate eigenvalues 
\begin{eqnarray}
E_{\pm} &=& \frac{1}{2} ( \mathrm{x}^1+ \mathrm{x}^2 \notag \\
&\pm& \sqrt{(\mathrm{x}^1)^2 + (\mathrm{x}^2)^2 + 2(2|\langle
  \mathbf{a}|\mathbf{b}\rangle|^2-1)\mathrm{x}^1 \mathrm{x}^2} ),
\label{H-spectrum}
\end{eqnarray}
and the $(N-2)$-fold degenerate eigenvalue 
\begin{eqnarray}
 E_>=\mathrm{x}^1+ \mathrm{x}^2.
\end{eqnarray}
Thus, the gap between the ground state ($E_-$) and the first excited state ($E_+$) becomes
\begin{eqnarray}
& \Delta(\mathbf{x}) = \sqrt{(\mathrm{x}^1)^2+ (\mathrm{x}^2)^2 + 2(2|\langle\mathbf{a} |\mathbf{b}\rangle |^2-1)\mathrm{x}^1 \mathrm{x}^2}.
\label{gap}
\end{eqnarray}

The Hamiltonian (\ref{H-matrixform}) can be diagonalized by noting that one can rewrite
\begin{eqnarray}
 H(\mathbf{x}) &=& \frac{1}{2}A(\mathbf{x})[\Delta(\mathbf{x})\Sigma_z -(\mathrm{x}^1+\mathrm{x}^2)I_{\{0,1\}}]A^{\dag}(\mathbf{x}) \nonumber\\
& & + (\mathrm{x}^1 + \mathrm{x}^2)I,
\label{H-matrixform-2}
\end{eqnarray}
where $\Sigma_z$ is the Pauli matrix
$\sigma_z=\mathrm{diag}(1,-1)\equiv|0\rangle\langle 0|-|1\rangle
\langle 1|$ padded with zeros to embed it trivially into the
$N$-dimensional representation (i.e.,
$\Sigma_z=\mathrm{diag}(\sigma_z,0,\ldots,0)$),
$I_{\{0,1\}}\equiv\mathrm{diag}(1,1,,0,\ldots,0)$, and the $2\times 2$
unitary matrix $A(\mathbf{x})$ is defined as
\begin{eqnarray}
 A(\mathbf{x})=e^{-i\theta(\mathbf{x})\sigma_y},
\end{eqnarray}
(the extension to $N$ dimensions is similar to that of $\Sigma_z$ by
padding with sufficiently many zeros)
with
\begin{eqnarray}
 &&\hskip-3mm\cos\theta = 2\mathrm{x}^2|\langle
  \mathbf{a}|\mathbf{b}\rangle|\sqrt{1-|\langle
    \mathbf{a}|\mathbf{b}\rangle|^2}/ \bigl\{4|\langle
  \mathbf{a}|\mathbf{b}\rangle|^2(1-|\langle
  \mathbf{a}|\mathbf{b}\rangle|^2)\nonumber\\ 
&&~\hskip-3mm \times(\mathrm{x}^2)^2 +\bigl( \mathrm{x}^1-[1-2|\langle \mathbf{a}|\mathbf{b}\rangle|^2]\mathrm{x}^2-\Delta \bigr)^2\bigr\}^{-1/2}.
\end{eqnarray}
After removing the energy shift
$(\mathrm{x}^1+\mathrm{x}^2)I$ from Eq.~(\ref{H-matrixform-2}), it is evident that the ground-state projection is 
\begin{eqnarray}
 P_0(\mathbf{x})=A(\mathbf{x})|1\rangle \langle 1|A^{\dag}(\mathbf{x}),
\end{eqnarray}
(padded with zeros). This yields
\begin{eqnarray}
 \mathbf{g}_{ij} &=&\mathrm{Tr}[\partial_i P_0\partial_j P_0]\nonumber\\
& = & -\partial_i\theta \partial_j \theta~ \mathrm{Tr}\bigl( [\sigma_y,P_0]^2\bigr)\nonumber\\
& = & \partial_i\theta \partial_j \theta.
\end{eqnarray}

Obtaining the geodesic for the one-dimensional case
$\mathbf{x}=(1-\mathrm{x},\mathrm{x})$ turns out to be simple
and can be performed analytically, yielding 
\begin{eqnarray}
\mathrm{x}(s)=\frac{1}{2}-\frac{|\langle \mathbf{a}|\mathbf{b}%
\rangle|}{2\sqrt{1-|\langle \mathbf{a}|\mathbf{b}\rangle|^2}}\tan
[(1-2s)\arccos|\langle \mathbf{a}|\mathbf{b}\rangle|].\nonumber\\
\end{eqnarray}
It is interesting to note that this is exactly the solution obtained
in Ref.~\cite{QAB} from the different metric $\widetilde{\mathbf{g}}$ [Eq.~(\ref{gtilde})].

\subsubsection{One-dimensional transverse-field Ising chain}
\label{ISING-EXAMPLE}

Consider a one-dimensional chain of spin-$1/2$ particles interacting
according to the following Hamiltonian:
\begin{eqnarray}
 H\bigl(\mathbf{x}(s) \bigr) = -\sum_{\ell=-m}^m \mathrm{x}^1(s)
\sigma_z^{(\ell)} + \mathrm{x}^2(s) \sigma_x^{(\ell)} \sigma_x^{(\ell+1)},
\label{ising-tr}
\end{eqnarray}
with the boundary conditions $\mathbf{x}_0=(1,0)$, $\mathbf{x}_1=(0,1)$, and 
$\bm{\sigma}^{(m+1)}\equiv \bm{\sigma}^{(1)}$ \cite{Schaller}. Exact
diagonalization by the Jordan-Wigner transformation \cite{Sachdev-book}
yields 
\begin{eqnarray}
 |\Phi_0(\mathbf{x})\rangle = \otimes_{\ell=1}^{m}\bigl( \cos\theta_{\ell}(\mathbf{x}) |0\rangle_{-\ell} |0\rangle_{\ell} + i \sin\theta_{\ell}(\mathbf{x}) |1\rangle_{-\ell} |1\rangle_{\ell} \bigr),\nonumber\\
 \label{GS-1}
\end{eqnarray}
where (cf. Ref.~\cite{Zanardi-prl:07})
\begin{eqnarray}
\hskip-2mm\sin 2\theta_{\ell} = \frac{\mathrm{x}^2\sin(\frac{2\pi \ell}{2m+1})}{\sqrt{(
\mathrm{x}^2\cos \frac{2\pi \ell}{2m+1} - \mathrm{x}^1)^2 + (\mathrm{x}^2)^2
\sin^2 \frac{2\pi \ell}{2m+1}}}.
\end{eqnarray}
It is evident from Eq.~(\ref{GS-1}) that
\begin{eqnarray*}
 |\dot{\Phi}_0\rangle & =& \sum_{i=1}^2  \dot{\mathrm{x}}^{i} \partial_{i}|\Phi_0 \rangle  \nonumber\\
& =& \sum_{i=1}^2  \dot{\mathrm{x}}^i\sum_{\ell=1}^m \partial_{i} \theta_{\ell} \bigl( -\sin\theta_{\ell} |0\rangle_{-\ell} |0\rangle_{\ell} + i \cos\theta_{\ell} |1\rangle_{-\ell} |1\rangle_{\ell} \bigr)\nonumber\\
&&~ \otimes|\Phi_{\bar{\ell}}\rangle,
\end{eqnarray*}
where $ |\Phi_{\bar{\ell}}\rangle$ is the same as $|\Phi_0\rangle$
[Eq.~(\ref{GS-1})] except that the term with the
label $\ell$ is absent. In addition it is easily verified that
$\langle \Phi_0|\dot{\Phi}_0\rangle=0$. Thus, we obtain
\begin{eqnarray}
 \langle \dot{\Phi}_0|\dot{\Phi}_0 \rangle = \sum_{i,j=1}^2 \dot{\mathrm{x}}^{i} \dot{\mathrm{x}}^{j}\sum_{\ell}^m  \partial_{i}\theta_{\ell}\partial_{j}\theta_{\ell}.
\end{eqnarray}
After inserting these results into Eq.~(\ref{g2}) we have
\begin{eqnarray}
 \mathbf{g}_{ij}(\mathbf{x})= \sum_{\ell=1}^m \partial_{i}\theta_{\ell}(%
\mathbf{x}) \partial_{j} \theta_{\ell}(\mathbf{x}).  \label{g-ising}
\end{eqnarray}
This, then, is the geometric tensor for the transverse field Ising model.

To make further progress we focus
on the one-parameter cases: (i) $\mathbf{x}%
=(1-\mathrm{x},\mathrm{x})$, (ii) $\mathbf{x}=(\mathrm{x},1)$, and (iii) $%
\mathbf{x}=(1,\mathrm{x})$, all subject to the boundary conditions $\mathrm{x}%
(0)=1-\mathrm{x}(1)=0$.

Let \begin{eqnarray}
p(\mathrm{x})=\frac{1}{4}\sum_{\ell=1}^m \frac{\sin^2(\frac{2\pi
    \ell}{2m+1})}{[1-2(1+\cos \frac{2\pi
      \ell}{2m+1})(1-\mathrm{x})\mathrm{x}]^2}.
\label{p(x)}
\end{eqnarray}
For a given finite lattice size $m$, the geodesic equation for case (i) reads
\begin{eqnarray}
2p(\mathrm{x}) \ddot{\mathrm{x}} + \partial_\mathrm{x} p(\mathrm{x})(\dot{\mathrm{x}})^2=0.
\label{eq-isi-1}
\end{eqnarray}
This equation can be integrated to yield
\begin{eqnarray}
 2s=\int_0^{\mathrm{x}(s)}\sqrt{p(\mathrm{x}%
^{\prime })}\mathrm{d}\mathrm{x}^{\prime }/ \int_0^{1/2}\sqrt{p(\mathrm{x}%
   ^{\prime })}\mathrm{d}\mathrm{x}^{\prime },
 \label{105}
\end{eqnarray}

We next consider the thermodynamic limit $m\rightarrow\infty$, where
we can obtain a simple closed-form formula for the geodesic.
The expression in this limit follows from substituting $\sum_{\ell}%
\rightarrow \frac{2m+1}{2\pi}\int_0^{\pi}\mathrm{d}\mathrm{z}$ [with $%
\mathrm{z}_{\ell}=2\pi \ell/(2m+1)$] and taking into account that the
model exhibits a QPT at $\mathrm{x}_c=1/2$ corresponding to $s_c=1/2$. This
yields
\begin{eqnarray}
  \mathrm{x}(s)=\begin{cases}
  \frac{1}{2}\bigl(1-\tan^2[\frac{\pi}{4}(1-2s)] \bigr),~~0\leq
  s\leq\frac{1}{2},  \label{g-I-a}\\
  \frac{1}{2}\bigl(1+\tan^2[\frac{\pi}{4}(1-2s)]
  \bigr),~~\frac{1}{2}\leq s\leq 1.  \label{g-I-2}
  \end{cases}
\end{eqnarray}
For details of the derivation see Appendix~\ref{app:x(s)}.

Similarly, for both cases (ii) and (iii) we obtain the geodesic for a given
finite $m$ as 
\begin{eqnarray}
 s=\int_0^{\mathrm{x}(s)}\sqrt{q(\mathrm{x}^{\prime })}\mathrm{%
d}\mathrm{x}^{\prime }/ \int_0^{1}\sqrt{q(\mathrm{x}^{\prime })}\mathrm{d}%
 \mathrm{x}^{\prime },
 \label{108}
\end{eqnarray}
where 
\begin{eqnarray}
q(\mathrm{x})=\frac{1}{4}\sum_{\ell=1}^m
\frac{\sin^2(\frac{2\pi \ell}{2m+1})}{[1-2\cos \frac{2\pi \ell}{2m+1}]^2}. 
\end{eqnarray}
In the thermodynamic limit a quantum critical point emerges at $\mathrm{x}_c=1$ ($%
s_c=1$), and a similar approach as in case (i) yields the geodesic 
\begin{eqnarray}
& \mathrm{x}(s)=\sin(\pi s/2).  
\label{g-I-bc}
\end{eqnarray}
For details of the derivation again see Appendix~\ref{app:x(s)}.

Figure~\ref{geo} illustrates the geodesics obtained
for the transverse field Ising model subject to the three
parametrizations we have discussed.
\begin{figure}[htp]
\includegraphics[width=4.2cm,height=3.2cm]{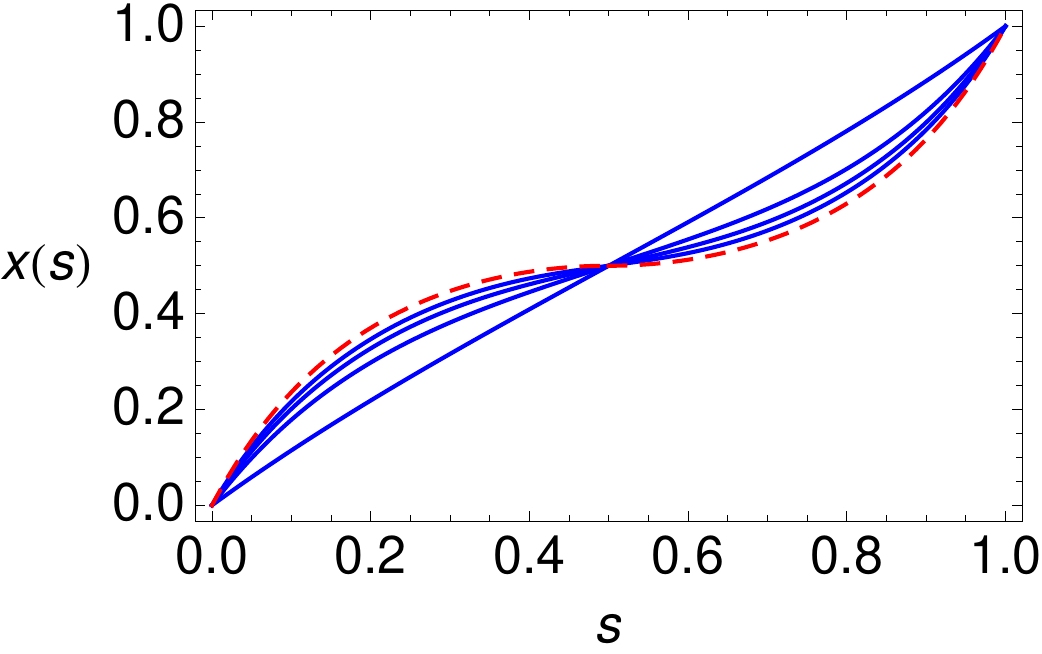} \hskip1mm %
\includegraphics[width=4.2cm,height=3.2cm]{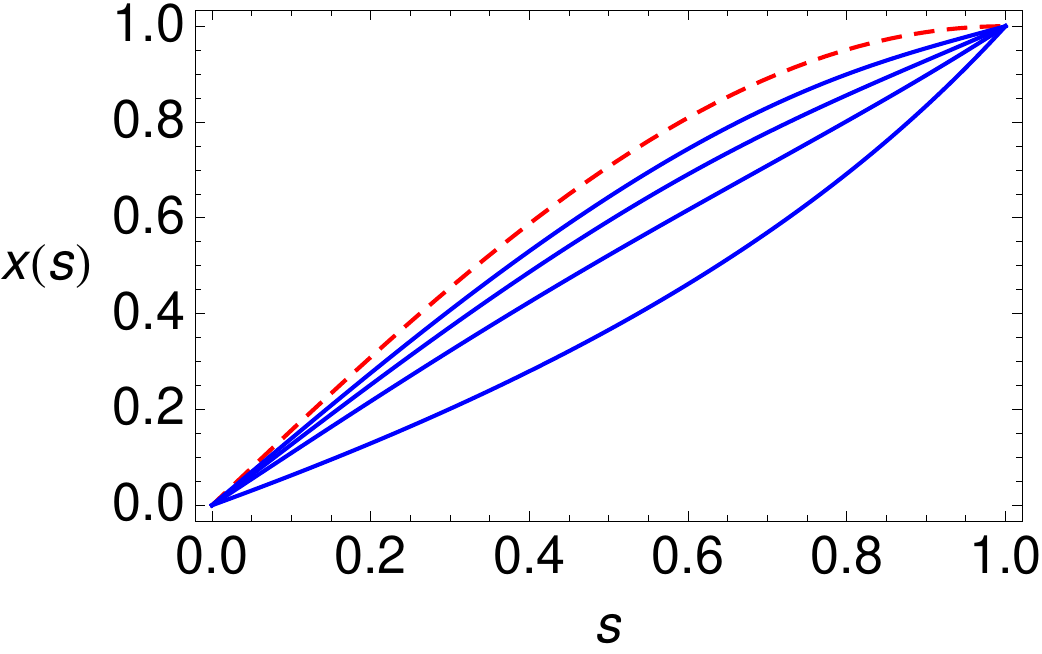}
\caption{(Color online) Optimal adiabatic paths for the one-dimensional
transverse-field Ising model, corresponding to the parameterizations $%
\mathbf{x}=(1-\mathrm{x},\mathrm{x})$ (left),
$\mathbf{x}=(\mathrm{x},1)$ and $\mathbf{x}=(1,\mathrm{x})$
(right). The red dashed lines represent the thermodynamic limit,
while the solid blue lines correspond to
$m=1,4,10,30,100$, approaching the dashed line as $m$ increases.}
\label{geo}
\end{figure}

%%%%%%%%%%%%%%%%%%%%%%%%%%%%%%%%%%%%%%%%%%%%%%%%%%%%%%%%%%%
\subsection{Geodesic for passage through a quantum critical point}
\label{geodesic}

A limitation of our formalism is that, in principle, exact knowledge
of the ground state is required in order to obtain the
geodesic. Unfortunately, such knowledge is rarely available, the
exceptions being certain exactly solvable models such as those we
treated in the previous subsection. With partial knowledge or an
approximation for the gap, one should solve Eq.~(\ref{EL-eq}) on a
case by case basis, possibly numerically.

However, while these observations apply in a setting where one wishes
to obtain the geodesic over the entire parameter manifold, the
situation in the vicinity of a quantum critical point is rather
different. Indeed, the most interesting physics usually happens in the
vicinity of the quantum criticality. In addition, the behavior of a quantum
adiabatic algorithm is essentially governed by how the system approaches
and/or passes through a quantum critical region. These considerations
suggest that knowledge of the geodesic around the quantum critical region should suffice for
most algorithmic or physically relevant applications, thus
obviating the need for knowing $P_0$ everywhere.

Computation of the critical behavior of other geometric functions, such as
$\Gamma$ and $\mathbf{R}$, is straightforward.
E.g., in the one-parameter case, where $\mathbf{x}=(\mathrm{x})$, the
Euler-Lagrange (geodesic) equation (\ref{EL-eq}) in the critical region slightly before and
after the critical point reduces to $\ddot{\mathrm{x}}+\nu\kappa\dot{\mathrm{x}}%
^2/2\mathrm{x}=0$, whence 
\begin{eqnarray}
& \mathrm{x}(s\approx s_c)\approx\mathrm{x}_c+ A(s-s_c)^{\chi}.
\label{ad-passage}
\end{eqnarray}
After using $\alpha=d+z-1/\nu$ \cite{Alet:09}, 
where $\alpha$ is the scaling dimension [recall Eq.~(\ref{crit-dim})],
we obtain
\begin{eqnarray}
  \chi=2/(2+\nu\kappa)=2/d\nu>0,
\label{eq:chi}
\end{eqnarray}
with $A$ constant (derivation details are given in Appendix~\ref{app:x(s)}).
This is a remarkable result as
it characterizes the optimal adiabatic passage through a quantum critical
point in terms of the universality class of the system. Moreover, this
result confirms that the critical geodesic has a power-law dependence
on $s$, as first reported in Ref.~\cite{others1}, although away from
the critical region the dependence can be different.
References \cite{others1,others2,others3,others4} report critical behaviors of the metric tensor and
related parameters obtained using different methods, such as
minimizing exact expressions for transition probability in
thermodynamic limit. In contrast to the result of Ref.~\cite{others1},
in our analysis the exponent $\chi$ of the critical geodesic depends on the dimensionality $d$,
whereas it is independent of the total time $T$. In adiabatic
evolution the dependence on $T$ is of course expected; however, note
that our scaling result depends only upon the geometry of the control
manifold, which does not depend on $T$.

%%%%%%%%%%%%%%%%%%%%%%%%%%%%%%%%%%%%%%%%%%%%%%%%%%%%%%%%%%%

\section{Summary and Conclusions}
\label{Summ}

In this work we set out to elucidate the role of geometry in adiabatic
quantum evolution. By splitting the ``adiabatic error'', i.e., the
norm of the difference between the ideal adiabatic evolution operator
and the actual propagator, into two components, one of which is
endowed with a geometric meaning, we were able to derive a Riemannian
metric tensor which encodes the geometry of adiabatic evolution. This
metric is capable of describing evolution over both nondegenerate and degenerate subspaces. We
then showed that this same metric tensor arises naturally from a number of
different but complementary viewpoints, including a minimization of
the operator fidelity, and a focus on the Grassmannian structure of the
dynamics.

Our second major goal in this work was to establish a firm
connection between adiabatic evolution and quantum phase
transitions. By analyzing the infinitesimal variation in the operator
fidelity we showed that, in fact the same metric tensor, arises in
both cases. We further derived the quantum critical scaling of this
metric tensor.

Having established a unified geometric framework for adiabatic quantum
evolution and quantum phase transitions, we proceeded to find the
geodesics on the manifold described by the unifying Riemannian metric
tensor. Such geodesics are of particular interest in adiabatic quantum
computing, where they correspond to paths which minimize the geometric
component of the deviation between the actual and desired final
states. We analytically determined the geodesics in three examples of
interest: the Deutsch-Jozsa algorithm, a generalization of Grover's
algorithm, and a model described by the transverse field Ising
model. While such examples are important as proofs of principle, one
cannot in general hope to analytically find the geodesics. For this
reason we focused on the passage through the quantum critical point,
and showed that in general, for second order QPTs, the geodesic in
this case obeys a universal scaling relation.

Among other applications, we expect that the formalism we have developed will lead to further
developments in adiabatic quantum computing, where the role of
criticality is well appreciated. We expect additional applications in
holonomic quantum computing, where degeneracy plays an essential role,
and where a differential geometric analysis of gate error minimization
has not yet been carried out.

%%%%%%%%%%%%%%%%%%%%%%%%%%%%%%%%%%%%%%%%%%%%%%%%%%%%%%%%%%%
\section{Acknowledgments}

Supported by NSF under Grants No.~PHY-802678 and CCF-726439 (to D.A.L.),
and PHY-803304 (to P.Z. and D.A.L.). D.F.A. acknowledges support by
a John Stauffer fellowship from the University of Southern California.\\

\noindent \textit{Note added}.---While this work was being finalized
for submission, a related manuscript appeared \cite{Ma:10}, which
similarly proposes a generalized quantum geometric tensor related to
adiabatic evolution of quantum many-body systems. 

%%%%%%%%%%%%%%%%%%%%%%%%%%%%%%%%%%%%%%%%%%%%%%%%%%%%%%%%%%%
%%%%%%%%%%%%%%%%%%%%%%%%%%%%%%%%%%%%%%%%%%%%%%%%%%%%%%%%%%%
\begin{widetext}
\appendix{}

%%%%%%%%%%%%%%%%%%%%%%%%%%%%%%%%%%%%%%%%%%%%%%%%%%%%%%%%%%%
\section{Proof of the Wilczek-Zee holonomy formula}
\label{app:WZformula-proof}

Notice that from the fact that $P_{0}(s)$ is a projector, i.e., $%
P_{0}(s)=P_{0}^{2}(s)$, we obtain 
\begin{equation}
\dot{P}_{0}(s)=\dot{P}_{0}(s)P_{0}(s)+P_{0}(s)\dot{P}_{0}(s),  \label{P0}
\end{equation}%
(where $\dot{P}_{0}(s)\equiv \partial _{s}P_{0}(s)$), so that 
\begin{equation}
P_{0}(s)\dot{P}_{0}(s)P_{0}(s)=0,  \label{PdPP}
\end{equation}%
and
\begin{equation}
[\dot{P}_0(s),P_0(s)] = 2\dot{P}_0(s)P_0(s) -\dot{P}_0(s).
\end{equation}

Let $Q_{0}(s)$ denote the projector orthogonal to $P_{0}(s)$, i.e., $P_{0}(s)+Q_{0}(s)=I$. Then we have
\begin{equation}
P_{0}(s)Q_{0}(s)=Q_{0}(s)P_{0}(s)=0. 
\label{PQ}
\end{equation}%

The differential equation for $V_{\alpha \alpha^{\prime } }^{[0]}(s)$ [Eq.~(\ref{WZFORMULA})] can be obtained as follows:
\begin{equation}\label{eq1}
\partial_sV_{\alpha  \alpha^{\prime }}^{[0]}(s) = \langle \dot{\Phi} _{0}^{\alpha }(s)|V_{\text{ad}}(s)|\Phi _{0}^{\alpha^{\prime} }(0)\rangle + \langle \Phi _{0}^{\alpha }(s)|\dot{V}_{\text{ad}}(s)|\Phi _{0}^{\alpha^{\prime} }(0)\rangle.
\end{equation}
In addition, consider the action of $H_\mathrm{ad}(s)$
[Eq.~(\ref{Had-H})] on
$|\Phi _{0}^{\alpha }(s)\rangle$, 
\begin{eqnarray}\label{eq2}
H_\mathrm{ad}(s)|\Phi _{0}^{\alpha }(s)\rangle &=& \left(H(s) + 2i\dot{P}_0(s)P_0(s)/T - i\dot{P}_0(s)/T\right)|\Phi _{0}^{\alpha }(s)\rangle \nonumber \\
&=& E_0(s)|\Phi _{0}^{\alpha }(s)\rangle + i \dot{P}_0(s)|\Phi _{0}^{\alpha }(s)\rangle/T.
\end{eqnarray}
Since $\dot{P}_{0}(s)=\sum_{\beta=1}^{g_0} |\dot{\Phi}_{0}^{\beta }(s)\rangle \langle \Phi_{0}^{\beta }(s)|+|\Phi _{0}^{\beta }(s)\rangle \langle \dot{\Phi}_{0}^{\beta }(s)| $, 
we have
\begin{equation}\label{eq3}
\dot{P}_0(s)|\Phi _{0}^{\alpha }(s)\rangle =  |\dot{\Phi}_{0}^{\alpha }(s)\rangle + \sum_{\beta=1}^{g_0} \langle\dot{\Phi}_{0}^{\beta}(s)|\Phi _{0}^{\alpha }(s)\rangle |\Phi_{0}^{\beta}(s)\rangle.
\end{equation}
Using Eq.~(\ref{eq2}) and (\ref{eq3}), we can rewrite Eq.~(\ref{eq1}) as
\begin{eqnarray}
  \partial_sV_{\alpha  \alpha^{\prime }}^{[0]}(s)
&=&  \langle \dot{\Phi} _{0}^{\alpha }(s)|V_{\text{ad}}(s)|\Phi _{0}^{\alpha ^{\prime}}(0)\rangle  -iT \langle \Phi _{0}^{\alpha }(s)|H_\text{ad}(s) V_{\text{ad}}(s)|\Phi _{0}^{\alpha^{\prime} }(0)\rangle \nonumber\\
&=&  - iT E_0(s) \langle\Phi _{0}^{\alpha }(s)| V_{\text{ad}}(s) |\Phi _{0}^{\alpha^{\prime}}(0)\rangle - \sum_{\beta=1}^{g_0} \langle\Phi _{0}^{\alpha }(s) |\dot{\Phi} _{0}^{\beta}(s)\rangle  \langle\Phi _{0}^{\beta}(s)| V_{\text{ad}}(s) |\Phi _{0}^{\alpha^{\prime}}(0)\rangle\\                 \nonumber  
\end{eqnarray}
Without loss of generality, after setting $E_0(s) =0$, we obtain the following differential equation for $V_{\alpha\alpha ^{\prime }}^{[0]}(s)$:
\begin{eqnarray}
\partial_sV_{\alpha  \alpha^{\prime }}^{[0]}(s) &=& - \sum_{\beta=1}^{g_0} \langle\Phi _{0}^{\alpha }(s) |\dot{\Phi} _{0}^{\beta}(s)\rangle  \langle\Phi _{0}^{\beta}(s)| V_{\text{ad}}(s) |\Phi _{0}^{\alpha^{\prime}}(0)\rangle \nonumber 
\\
&=& - \sum_{\beta=1}^{g_0}A_{\alpha\beta}(s) V_{\beta\alpha^{\prime}}^{[0]}(s),
\end{eqnarray}
whose solution is 
\begin{equation}
V^{[0]}(s)=\mathcal{P}\exp \left( -\int_{0}^{s}A(s^{\prime })\mathrm{d}s^{\prime }\right),
\end{equation}%
with 
\begin{equation}
A_{\alpha \beta}\equiv \langle \Phi _{0}^{\alpha }|\partial
_{s}|\Phi _{0}^{\beta}\rangle.
\end{equation}

%%%%%%%%%%%%%%%%%%%%%%%%%%%%%%%%%%%%%%%%%%%%%%%%%%%%%%%%%%%
\section{Proof of E\lowercase{q}.~(\protect\ref{PDOTP-NORM})}
\label{app:PdotP-norm-proof}

Equation~(\ref{PdPP}) yields
\begin{equation}
\lbrack \dot{P}_{0},P_{0}]^{2}=-(\dot{P}_{0}P_{0}\dot{P}_{0}+P_{0}\dot{P}%
_{0}^{2}P_{0}).  \label{P0comm}
\end{equation}%
Using Eq.~(\ref{P0}) to write $P_{0}\dot{P}_{0}=\dot{P}_{0}-\dot{P}_{0}P_{0}$
and substituting this into the first term of Eq.~(\ref{P0comm}) we then have%
\begin{eqnarray}
\lbrack \dot{P}_{0},P_{0}]^{2} &=&-(\dot{P}_{0}^{2}-\dot{P}%
_{0}^{2}P_{0}+P_{0}\dot{P}_{0}^{2}P_{0})  \notag \\
&=&-\dot{P}_{0}^{2}+Q_{0}\dot{P}_{0}^{2}P_{0}.  \label{P0comm2}
\end{eqnarray}%
The second term vanishes, as can be seen by using Eq.~(\ref{P0}) to write $%
\dot{P}_{0}^{2}=(\dot{P}_{0}P_{0}+P_{0}\dot{P}_{0})^{2}$:%
\begin{equation*}
Q_{0}\dot{P}_{0}^{2}P_{0}=Q_{0}(\dot{P}_{0}P_{0}\dot{P}_{0}P_{0}+\dot{P}%
_{0}P_{0}\dot{P}_{0}+P_{0}\dot{P}_{0}^{2}P_{0}+P_{0}\dot{P}_{0}P_{0}\dot{P}%
_{0})P_{0}=0,
\end{equation*}%
where we used Eq.~(\ref{PdPP}) on the first two summands and Eq.~(\ref{PQ})
on the last two. Thus we conclude that 
\begin{equation}
\lbrack \dot{P}_{0},P_{0}]^{2}=-\dot{P}_{0}^{2}.  \label{temp}
\end{equation}

Note that $\dot{P}_{0}=\sum_{\alpha=1}^{g_0} |\dot{\Phi}_{0}^{\alpha }\rangle \langle \Phi
_{0}^{\alpha }|+|\Phi _{0}^{\alpha }\rangle \langle \dot{\Phi}_{0}^{\alpha
}| $ is Hermitian and that therefore $[\dot{P}_{0},P_{0}]$ is
anti-Hermitian. Thus both $\dot{P}_{0}$ and $[\dot{P}_{0},P_{0}]$ are
unitarily diagonalizable: $-\dot{P}_{0}=VDV^{\dag }$, $[\dot{P}%
_{0},P_{0}]=WEW^{\dag }$, where $V$ and $W$ are unitary, while $D$ and $E$ are the
diagonal matrices of eigenvalues. Therefore it follows from Eq.~(\ref%
{P0comm2}) that $\Vert VD^{2}V^{\dag }\Vert =\Vert WE^{2}W^{\dag }\Vert $,
and from the unitary invariance of the operator norm that $\Vert D^{2}\Vert =\Vert E^{2}\Vert $. From
here we conclude that the maximum absolute values of their eigenvalues are
equal, i.e.: 
\begin{equation}
\Vert \lbrack \dot{P}_{0},P_{0}]\Vert =\Vert \dot{P}_{0}\Vert .
\end{equation}%
It also follows that $\Vert \dot{P}_{0}^{2}\Vert =\Vert \lbrack \dot{P}%
_{0},P_{0}]^{2}\Vert =\Vert D^{2}\Vert =\Vert D\Vert ^{2}=\Vert \lbrack \dot{%
P}_{0},P_{0}]\Vert ^{2}$, i.e., 
\begin{equation}
\Vert \lbrack \dot{P}_{0},P_{0}]\Vert =\sqrt{\Vert \dot{P}_{0}^{2}\Vert }.
\label{temp2}
\end{equation}

Next we wish to show that 
\begin{equation}
\dot{P}_{0}=-\left( P_{0}\dot{H}\frac{1}{H-E_{0}}+\frac{1}{H-E_{0}}\dot{H}%
P_{0}\right).  \label{pdot}
\end{equation}
To prove this note first that the Hamiltonian can be decomposed as 
\begin{equation}
H=E_{0}P_{0}+Q_{0}HQ_{0}.
\end{equation}%
Then 
\begin{equation}
\dot{H}=\dot{E}_{0}P_{0}+E_{0}\dot{P}_{0}-\dot{P}_{0}HQ_{0}+Q_{0}\dot{H}%
Q_{0}-Q_{0}H\dot{P}_{0},  \label{E*}
\end{equation}%
and multiplying this equation by $P_{0}$ from the right while using Eqs.~(%
\ref{PdPP}) and (\ref{PQ}) and the fact that $H$ commutes with $P_{0}$,
yields%
\begin{eqnarray}
\dot{H}P_{0} & =&\dot{E}_{0}P_{0}+E_{0}\dot{P}_{0}P_{0}-(I-P_{0})H\dot{P}%
_{0}P_{0}  \notag \\
&=&\dot{E}_{0}P_{0}+E_{0}\dot{P}_{0}P_{0}-H\dot{P}_{0}P_{0}.
\label{bbb}
\end{eqnarray}%
The operator $H-E_{0}I$ is invertible when its domain excludes the spectrum
of $H$ (and is then called the ``reduced resolvent;" see, e.g., Ref.~\cite{LRH}). 
That is, the inverse is defined as $Q_0[H-E_0]^{-1}Q_0$ (but for
brevity and when there is no risk of confusion, we simply write
$[H-E_0]^{-1}$ henceforth). 
With this restriction in mind we then have 
\begin{equation}
\dot{P}_{0}P_{0}=-\frac{1}{H-E_{0}}(\dot{H}-\dot{E}_{0})P_{0}=-\frac{1}{%
H-E_{0}}\dot{H}P_{0},  \label{PdotP}
\end{equation}%
where in the last step we used 
\begin{equation}
\frac{1}{H-E_{0}}P_{0}=P_{0}\frac{1}{H-E_{0}}=0,  \label{res}
\end{equation}%
which is due to the fact that the range of $[H-E_{0}]^{-1}$ is the range
of $Q_{0}$ [recall also Eq.~(\ref{resolvent})]. Similarly, by multiplying
Eq.~($\text{\ref{E*}) from the left by }P_{0}$ we obtain:%
\begin{equation}
P_{0}\dot{P}_{0}=-P_{0}\dot{H}\frac{1}{H-E_{0}}.  \label{Eb}
\end{equation}%
Adding Eqs.~(\ref{PdotP}) 
and (\ref{Eb}), and using Eq.~(\ref{P0}) again then
yields Eq.~(\ref{pdot}).

As a corollary, we can also calculate $\dot{E}_0(s)$ from Eq.~(\ref{bbb})
\begin{eqnarray}
 \dot{E}_0(s) = \mathrm{Tr}[\dot{H}P_0]/g_0.
\label{Edot}
\end{eqnarray}

Calculation of $\ddot{P}_0$ or higher order derivatives of $P_0$
follows similar logic (see, for example, Ref.~\cite{LRH}). For
example, we obtain 
\begin{eqnarray}
 \ddot{P}_0=-\Big( \dot{P}_0\dot{H}\frac{1}{H-E_0}+P_0 \ddot{H}\frac{1}{H-E_0} + P_0\dot{H}\partial_s\left[\frac{1}{H-E_0} \right] + \partial_s\left[\frac{1}{H-E_0} \right] \dot{H}P_0 + \frac{1}{H-E_0}\ddot{H} P_0 + \frac{1}{H-E_0}\dot{H} \dot{P}_0\Big).\nonumber\\
\label{ddp}.
\end{eqnarray}
This relation can be simplified further after replacing $\dot{P}_0$ [Eq.~(\ref{pdot})], using the identity
\begin{eqnarray}
 \partial_s\left[ \frac{1}{H-E_0}\right] =-\frac{1}{H-E_0}(\dot{H}-\dot{E}_0)\frac{1}{H-E_0},
\end{eqnarray}
and inserting $\dot{E}_0$ [Eq.~(\ref{Edot})]. However, we do not need the final explicit form here.

We are now ready to prove Eq.~(\ref{PDOTP-NORM}). Let 
\begin{equation}
A\equiv \frac{1}{H-E_{0}}\dot{H}P_{0},\quad B\equiv P_{0}\dot{H}\frac{1}{%
H-E_{0}}.
\end{equation}
Then, using Eqs.~(\ref{temp2}),~(\ref{pdot}), and (\ref{res}) yields

\begin{equation}
\Vert \lbrack \dot{P}_{0},P_{0}]\Vert =\sqrt{\Vert A^{\dag }A+B^{\dag
}B\Vert }.
\end{equation}%
Note that $A^{\dag }A$ and $B^{\dag }B$ are both positive operators and that
they have orthogonal support. Therefore 
$\Vert A^{\dag }A+B^{\dag }B\Vert =\max\{\Vert A^{\dag }A\Vert,\Vert B^{\dag }B\Vert\} $. 
Moreover, we have $A^{\dag
}A=BB^{\dag }$, and it is a basic property of the operator norm that $\Vert
BB^{\dag }\Vert =\Vert B^{\dag }B\Vert $ for any operator $B$. Thus $\sqrt{%
\Vert A^{\dag }A+B^{\dag }B\Vert }=\sqrt{\Vert A^{\dag }A\Vert }$, 
which is Eq.~(\ref{PDOTP-NORM}).

%%%%%%%%%%%%%%%%%%%%%%%%%%%%%%%%%%%%%%%%%%%%%%%%%%%%%%%%%%%
\section{Proof of the error formula in the Frobenius norm}
\label{app:Frob}

Starting from the definition of the adiabatic error, Eq.~(\ref{eps2}), we
have, by using Eq.~(\ref{PdPP}) together with $P_{0}^{2}=P_{0}$ and cyclic
invariance of the trace: 
\begin{eqnarray}
\epsilon (s) &=&\int_{0}^{s}\sqrt{\mathrm{Tr}[(P_{0}\dot{P}_{0}-\dot{P}%
_{0}P_{0})(\dot{P}_{0}P_{0}-P_{0}\dot{P}_{0})]}\mathrm{d}s^{\prime }  \notag
\\
&=&\int_{0}^{s}\sqrt{\mathrm{Tr}[P_{0}\dot{P}_{0}\dot{P}_{0}+\dot{P}_{0}P_{0}%
\dot{P}_{0}]}\mathrm{d}s^{\prime }  \notag \\
&=&\int_{0}^{s}\sqrt{\mathrm{Tr}[P_{0}(\partial _{i}P_{0})(\partial
_{j}P_{0})+(\partial _{i}P_{0})P_{0}(\partial _{j}P_{0})]\dot{\mathbf{x}}^{i}%
\dot{\mathbf{x}}^{j}}\mathrm{d}s^{\prime }\nonumber\\
\end{eqnarray}%
where $\dot{P}_{0}=\partial _{i}P_{0}\dot{\mathbf{x}%
}^{i}$. Using $P_{0}^{2}=P_{0}$ once more to obtain $P_{0}(\partial
_{i}P_{0})+(\partial _{i}P_{0})P_{0}=\partial _{i}P_{0}$ we have:%
\begin{eqnarray}
\epsilon (s) &=&\int_{0}^{s}\sqrt{\mathrm{Tr}[\{\partial _{i}P_{0}-(\partial
_{i}P_{0})P_{0}\}(\partial _{j}P_{0})+(\partial _{i}P_{0})P_{0}(\partial
_{j}P_{0})]\dot{\mathbf{x}}^{i}\dot{\mathbf{x}}^{j}}\mathrm{d}s^{\prime } 
\label{temp-1}
\\
&=&\int_{0}^{s}\sqrt{2g_0\mathbf{g}_{ij}(\mathbf{x})\dot{\mathbf{x}}^{i}\dot{%
\mathbf{x}}^{j}}\mathrm{d}s^{\prime }
\end{eqnarray}%
where the metric tensor is defined as $\mathbf{g}_{ij}\equiv \mathrm{Tr}%
[\partial _{i}P_{0}\partial _{j}P_{0}]/2g_0$, which is Eq.~(\ref{g2}).

Next let us derive Eq.~(\ref{metricH}). From Eq.~(\ref{pdot}) we have%
\begin{equation}
\partial _{i}P_{0}=-\left( P_{0}(\partial _{i}H)\frac{1}{H-E_{0}}+\frac{1}{%
H-E_{0}}(\partial _{i}H)P_{0}\right) .
\end{equation}%
Inserting this into $\mathrm{Tr}[\partial _{i}P_{0}\partial _{j}P_{0}]$ and
expanding the product while using Eq.~(\ref{res}), we obtain: 
\begin{eqnarray}
\mathrm{Tr}[\partial _{i}P_{0}\partial _{j}P_{0}] &=&\mathrm{Tr}\left[
\left( P_{0}(\partial _{i}H)\frac{1}{H-E_{0}}+\frac{1}{H-E_{0}}(\partial
_{i}H)P_{0}\right) \left( P_{0}(\partial _{j}H)\frac{1}{H-E_{0}}+\frac{1}{%
H-E_{0}}(\partial _{j}H)P_{0}\right) \right]  \notag \\
&=&\mathrm{Tr}\left[ P_{0}(\partial _{i}H)\left( \frac{1}{H-E_{0}}\right)
\left( \frac{1}{H-E_{0}}\right) (\partial _{j}H)P_{0}+\frac{1}{H-E_{0}}%
(\partial _{i}H)P_{0}P_{0}(\partial _{j}H)\frac{1}{H-E_{0}}\right]  \notag \\
&=&\mathrm{Tr}\left[ P_{0}(\partial _{i}H)\left( \frac{1}{H-E_{0}}\right)
^{2}(\partial _{j}H)P_{0}\right] +\mathrm{Tr}\left[ P_{0}(\partial
_{j}H)\left( \frac{1}{H-E_{0}}\right) ^{2}(\partial _{i}H)P_{0}\right] ,
\end{eqnarray}%
as desired.

%%%%%%%%%%%%%%%%%%%%%%%%%%%%%%%%%%%%%%%%%%%%%%%%%%%%%%%%%%%
\section{Proof that $\mathbf{g}$ is a metric}
\label{app:metric}

By definition, a metric must satisfy three properties \cite{Nakahara-book}:\
it must be positive, real, and symmetric.

(1) Positive:\ For any nonzero $\bm{\alpha }(\mathbf{x})\in T_{\mathcal{M%
}}(\mathbf{x})$ we have 
\begin{eqnarray}
\bm{\alpha }(\mathbf{x})\cdot \mathbf{g}(\mathbf{x})\cdot \bm{\alpha 
}(\mathbf{x}) &=&\mathbf{g}_{ij}(\mathbf{x})\bm{\alpha }^{i}(\mathbf{x})
\bm{\alpha }%
^{j}(\mathbf{x})\nonumber\\ &=&\frac{1}{2g_0}\mathrm{Tr}[\bigl(\partial _{i}P_{0}(\mathbf{x})\bigr)\bigl(\partial _{j}P_{0}(\mathbf{x})\bigr)]\bm{\alpha }^{i}(\mathbf{x})\bm{\alpha }^{j}(\mathbf{x})
\nonumber\\
 &=&
\mathrm{Tr}\left\{\left[\frac{1}{\sqrt{2g_0}}\bm{\alpha }^{i}(\mathbf{x})\partial _{i}P(\mathbf{x})\right]_{lk}\left[\frac{1}{\sqrt{2g_0}}
\bm{\alpha }^{j}(\mathbf{x})\partial _{j}P(\mathbf{x})\right]_{kl}\right\}
\nonumber\\ &\equiv& \mathrm{Tr}%
[C^{\dag }(\bm{\alpha },\mathbf{x})C(\bm{\alpha },\mathbf{x})]\geq 0,
\end{eqnarray}
where
\begin{eqnarray}
 C(\bm{\alpha },\mathbf{x}) &\equiv & \frac{1}{\sqrt{2g_0}}\bm{\alpha }^{i}(\mathbf{x}%
)\partial _{i}P(\mathbf{x}).
\end{eqnarray}

Note that although $\mathrm{Tr}[(\mathrm{d}P_0)^2]$ is always
positive, when we move to a coordinate $\mathbf{x}$ the resulting
\textit{pull-back} metric $\mathbf{g}(\mathbf{x})$ might become
singular (non-invertible) at some points or even identically zero. In
this strict sense
$\mathbf{g}(\mathbf{x})$ is not a metric.  

(2) Real: This is obvious from the very construction of $\mathbf{g}=\mathrm{Re}[\mathbf{G}]$.

(3) Symmetric:\ This is obvious from the definition and cyclic invariance of
the trace: $\mathbf{g}_{ij}\equiv \mathrm{Tr}[\partial _{i}P_{0}\partial
_{j}P_{0}]/2g_0=\mathrm{Tr}[\partial _{j}P_{0}\partial _{i}P_{0}]/2g_0=\mathbf{g}_{ji}$.

%%%%%%%%%%%%%%%%%%%%%%%%%%%%%%%%%%%%%%%%%%%%%%%%%%%%%%%%%%%
\section{Proof of the operator fidelity inequalities}
\label{app:op-fid}

We start by proving Eq.~(\ref{opfid-eps}). From the definition of the
operator fidelity, Eq.~(\ref{f}), with $\varrho =I/N$, we have, using
Eq.~(\ref{Om}):
\begin{eqnarray}
f(s) &=&\left\vert \mathrm{Tr}\left[ \frac{I}{N}\Omega (s)\right]
\right\vert =\left\vert \mathrm{Tr}\left[ \frac{I}{N}-\frac{1}{N}%
\int_{0}^{s}K_{T}(s^{\prime })\Omega (s^{\prime })\mathrm{d}s^{\prime }%
\right] \right\vert  \notag \\
&=&\left\vert 1-\frac{1}{N}\int_{0}^{s}\mathrm{Tr}[K_{T}\Omega ]\mathrm{d}%
s^{\prime }\right\vert  \notag \\
&\geq &1-\frac{1}{N}\int_{0}^{s}\left\vert \mathrm{Tr}[K_{T}\Omega
]\right\vert \mathrm{d}s^{\prime }  \notag \\
&=&1-\frac{1}{N}\int_{0}^{s}\left\vert \mathrm{Tr}([\partial _{s^{\prime
}}P_{0},P_{0}]V V_{\text{ad}}^{\dag })\right\vert \mathrm{d}s^{\prime },
\end{eqnarray}%
where in the last line we used the definitions of $\Omega (s)$
[Eq.~(\ref{Om-def})] and $K_{T}(s)$ [Eq.~(\ref{K-def})], and cyclic
invariance of the 
trace. Now recall the Cauchy-Schwartz inequality for operators \cite{Bhatia:book}: 
\begin{equation}
\Vert A\Vert _{2}\Vert B\Vert _{2}\geq \left\vert \langle A,B\rangle
\right\vert :=\left\vert \mathrm{Tr}[A^{\dag }B]\right\vert .
\end{equation}%
Applying this with $A:=[\partial _{s^{\prime }}P_{0},P_{0}]$ and 
$B:=V V_{\text{ad}}^{\dag }$ 
and noting that 
$\Vert V V_{\text{ad}}^{\dag }\Vert _{2}=
\sqrt{\mathrm{Tr}[V_{\text{ad}}V^{\dag }VV_{\text{ad}}^{\dag
}]}=\sqrt{N}$,
we obtain%
\begin{equation}
f(s)\geq 1-\frac{1}{\sqrt{N}}\int_{0}^{s}\Vert \lbrack \partial _{s^{\prime
}}P_{0},P_{0}]\Vert _{2}\mathrm{d}s^{\prime }=1-\frac{1}{\sqrt{N}}\epsilon
(s),
\end{equation}%
as we set out to prove. The inequality $f(s)\leq 1$ follows from the fact
that $\Omega (s)$ is unitary:\ diagonalizing $\Omega (s)$ and taking the
absolute values of all $N$ of its diagonal elements, which are roots of
unity, gives $\left\vert \mathrm{Tr}\left[ \Omega (s)\right] \right\vert
\leq N$.

Next we prove Eq.~(\ref{O-bound}). Using Eq.~(\ref{Om}) along with
submultiplicativity and the triangle inequality we have:%
\begin{eqnarray}
\Vert O(s)-O_{\text{ad}}(s)\Vert &=&\Vert O-\Omega (s)^{\dag }O\Omega
(s)\Vert  \notag \\
&=&\Vert O-(I-\sum_{l=1}^{\infty }\Omega _{l}^{\dag
}(s))O(I-\sum_{l'=1}^{\infty }\Omega _{l'}(s))\Vert  \notag \\
&=&\Vert \sum_{l=1}^{\infty }\Omega _{l}^{\dag }(s)O+O\sum_{l=1}^{\infty
}\Omega _{l}(s)-\sum_{l,l^{\prime }=1}^{\infty }\Omega _{l}^{\dag
}(s)O\Omega _{l^{\prime }}(s)\Vert  \notag \\
&\leq &\Vert O\Vert \sum_{l=1}^{\infty }\Vert \Omega _{l}(s)\Vert
(2+\sum_{l'=1}^{\infty }\Vert \Omega _{l'}(s)\Vert )  \notag \\
&=&\Vert O\Vert \left[ \Vert \Omega _{1}(s)\Vert +\sum_{l=2}^{\infty }\Vert
\int_{0}^{s}ds^{\prime }K_{T}(s^{\prime })\Omega _{l-1}(s^{\prime })\Vert %
\right] \left[ 2+\sum_{l'=1}^{\infty }\Vert \Omega _{l'}(s)\Vert \right] .
\end{eqnarray}%
The term in the first square brackets is identical to that in Eq.~(\ref%
{I-Om1}), and hence is bounded by Eq.~(\ref{I-Om2}). The summand $%
\sum_{l=1}^{\infty }\Vert \Omega _{l}(s)\Vert $ in the second term is $%
\mathcal{O}(1/T)$ according to Eq.~(\ref{Avron1}). 
We thus have:%
\begin{eqnarray}
\Vert O(s)-O_{\text{ad}}(s)\Vert &\leq &
\Vert O\Vert \left[\delta_1(s) +\widetilde{\epsilon}(s)\mathcal{O}(1/T)\right] \left[ 2+\mathcal{O}(1/T) \right],
\end{eqnarray}%
where $\delta _{1}$ 
is defined in Eq.~(\ref{del1}), 
and the last line follows from Eq.~(\ref{Avron1}).

%%%%%%%%%%%%%%%%%%%%%%%%%%%%%%%%%%%%%%%%%%%%%%%%%%%%%%%%%%%
\section{Proof of E\lowercase{q}.~(\ref{geometric tensor})}
\label{app:G-proof}

The operator fidelity of two positive operators $X$ and $Y$ relative to a density matrix $\varrho$ is defined as
\begin{eqnarray}
f_{\varrho}(X,Y)=\mathrm{Tr}[XY\varrho],
\end{eqnarray}
which is always nonnegative because the trace of the product of positive operators is nonnegative. 
When $X,Y\in\mathcal{G}_{N,g_0}$ [subsection \ref{grassmann}] and when
$\varrho$ is fully supported on the ground eigensubspace, one can conclude
from the inequality $0\leq\mathrm{Tr}[XY]\leq \mathrm{Tr}[Y]$
\cite{Bhatia:book} that $f_{\varrho}(X,Y)\leq1$.

Now we compute the fidelity of the ground-state projections $P_0(\mathbf{x})$ and $P_0(\mathbf{x}+\mathrm{d}\mathbf{x})$ relative to $\varrho=I_{g_0}/g_0$ up to the first nonvanishing order. Hence, 
\begin{eqnarray}
f_{\varrho}\bigl(P_0(\mathbf{x}),P_0(\mathbf{x}+\mathrm{d}\mathbf{x})\bigl) &=& \langle P_{0}(\mathbf{x}),P_{0}(\mathbf{x}+\mathrm{d}\mathbf{x})\rangle_{\varrho} \nonumber\\
& =& \frac{1}{g_0}\mathrm{Tr}[P_{0}(\mathbf{x})P_{0}(\mathbf{x}+\mathrm{d}\mathbf{x})]\nonumber\\
& =& \frac{1}{g_0}\mathrm{Tr}[P_{0}(\mathbf{x}) (P_0(\mathbf{x})+\mathrm{d}P_0(\mathbf{x})+\frac{1}{2}\mathrm{d}^2P_0(\mathbf{x}))]\nonumber\\
& = & 1 + \frac{1}{2g_0}\mathrm{Tr}[P_{0}(\mathbf{x})\mathrm{d}^2 P_0(\mathbf{x}) P_0(\mathbf{x})]
\label{geometrictensor}
\end{eqnarray}
where in the last two lines we used Eqs.~(\ref{P0}) and (\ref{PdPP}). Equation~(\ref{P0}) also yields
\begin{eqnarray}
\mathrm{d}^2P_0 = \mathrm{d}^2 P_0 P_0+ 2\mathrm{d}P_0 \mathrm{d}P_0 + P_0\mathrm{d}^2 P_0,
\end{eqnarray}
whence,
\begin{eqnarray}
P_0 \mathrm{d}^2 P_0 P_0 = -2P_0 (\mathrm{d}P_0)^2 P_0.
\end{eqnarray}
Thus Eq.~(\ref{geometrictensor}) is simplified as follows:
\begin{eqnarray}
f_{\varrho}\bigl(P_0(\mathbf{x}),P_0(\mathbf{x}+\mathrm{d}\mathbf{x})\bigl) &=& 1- \frac{1}{g_0}\mathrm{Tr}[P_{0} (\mathrm{d} P_0)^2 P_0].
\end{eqnarray}

%%%%%%%%%%%%%%%%%%%%%%%%%%%%%%%%%%%%%%%%%%%%%%%%%%%%%%%%%%%
\section{Proof of E\lowercase{qs}.~(\ref{g-bnd1}) and (\ref{g-bnd2})}
\label{app:g-bounds}

To prove Eq.~(\ref{g-bnd1}), we invoke the following inequality
\begin{eqnarray}
|\mathrm{Tr}[XY]|\leq \Vert X\Vert_1 \Vert Y\Vert,
\label{trineq}
\end{eqnarray}
valid for any pair of arbitrary operators $X$ and $Y$ \cite{ineq-2}.
In addition, note that by definition, the operator norm of the reduced resolvent $[H(s)-E_0(s)]^{-1}$ satisfies
\begin{eqnarray}
 \left\Vert \frac{1}{H(s)-E_0(s)}\right\Vert & =
 &\frac{1}{\mathrm{dist}\bigl(\{E_0(s)\},\mathrm{spec}\bigl( H(s)
   \bigr)\backslash \{E_0(s)\} \bigr)} \notag \\
& \leq & \frac{1}{\min_s \Delta(s)},
\label{norm-res}
\end{eqnarray}
where
$\mathrm{spec}\bigl(H(s) \bigr)$ is the spectrum of $H(s)$ and the
distance between two sets $\mathcal{A}$ and $\mathcal{B}$ is defined as follows:
\begin{eqnarray}
 \mathrm{dist}(\mathcal{A},\mathcal{B})\equiv \inf_{a\in\mathcal{A},b\in\mathcal{B}}|a-b|.
\end{eqnarray}

Equation~(\ref{metricH}) now yields:
\begin{eqnarray}
 \mathbf{g}_{ij} &\leq& |\mathbf{g}_{ij}|\nonumber\\
& \leq & \frac{1}{g_0}\left|\mathrm{Tr}\left[ (\partial _{j}H) P_{0}(\partial _{i}H)\left( \frac{1}{H-E_{0}}\right)
^{2} \right] \right|\nonumber\\
& \overset{\text{Eq.~(\ref{trineq})}}{\leq} &  \frac{1}{g_0} \Vert  (\partial _{j}H) P_{0}(\partial _{i}H)\Vert_1 \left\Vert \frac{1}{H-E_{0}} \right\Vert^2 \nonumber\\
& \overset{\text{Eq.~(\ref{norm-res}), submultiplicativity}}{ \leq} & \frac{1}{g_0~ \min_s \Delta^2}\Vert P_0\Vert_1 \Vert \partial_i H \partial_j H\Vert_1\nonumber\\
& \overset{\Vert P_0\Vert_1=g_0}{\leq} & \frac{\Vert \partial_i H \partial_j H\Vert_1}{\min_s \Delta^2}.
\end{eqnarray}

The proof of Eq.~(\ref{g-bnd2}) is immediate from noting that
$|\mathrm{Tr}[X]| \leq \sum_{i}\sigma_i(X)=\Vert X\Vert_1$. 

%%%%%%%%%%%%%%%%%%%%%%%%%%%%%%%%%%%%%%%%%%%%%%%%%%%%%%%%%%%%%%%%
\section{Proof of E\lowercase{q}.~(\ref{G-integral})}
\label{app:g-integral}

Note the following identity for the reduced resolvent:
\begin{eqnarray}
Q_0\frac{1}{H-E_0}Q_0 = Q_0\frac{1}{p+H-E_0}\Big |_{p=0}Q_0 = \int_0^{\infty} Q_0 e^{(-p + H - E_0)\tau} Q_0 ~\mathrm{d}\tau\Big |_{p=0}.
\label{app-red1}
\end{eqnarray}
Therefore
\begin{eqnarray}
\Big(Q_0\frac{1}{H-E_0}Q_0\Big)^2 =- \frac{\mathrm{d}}{\mathrm{d}p}Q_0 \frac{1}{p+H-E_0}\Big |_{p=0}Q_0
=- \frac{\mathrm{d}}{\mathrm{d}p} \int_0^{\infty} Q_0 e^{(-p + H -
  E_0)\tau} Q_0 \mathrm{d}\tau\Big |_{p=0}.
\label{app-red2}
\end{eqnarray}
Substituting
Eq.~(\ref{app-red2}) into Eq.~(\ref{eq-**}),
while recalling that in Eq.~(\ref{eq-**}) the inverse $[H-E_0]^{-1}$
is really shorthand for $Q_0[H-E_0]^{-1}Q_0$,
yields
\begin{eqnarray}
\mathbf{G}_{ij}&=&-\frac{1}{g_0}\frac{\mathrm{d}}{\mathrm{d}p}\int_0^{\infty} \mathrm{d}\tau~\mathrm{Tr}[P_0 (\partial_i H)  Q_0 e^{-(p + H-E_0)\tau}Q_0 (\partial_jH)]\Big |_{p=0}\nonumber\\
&=&-\frac{1}{g_0}\frac{\mathrm{d}}{\mathrm{d}p}\int_0^{\infty}\mathrm{d}\tau~
e^{-p\tau}\mathrm{Tr}[P_0 (\partial_i H_{\tau}) Q_0(\partial_j H)]\Big
|_{p=0}\nonumber \\ 
&=&-\frac{1}{g_0}\frac{\mathrm{d}}{\mathrm{d}p}\int_0^\infty
\mathrm{d}\tau~e^{-p \tau}\Big( \mathrm{Tr}[P_0 (\partial_i
  H_{\tau})(\partial_j H)] - \mathrm{Tr}[P_0 (\partial_i H) P_0
  (\partial_j H)]\Big)\Big |_{p=0}\label{app-Gij},
\end{eqnarray}
with $\partial_i H_{\tau} \equiv e^{\tau H}\partial_i H e^{-\tau
  H}$. Note that from Eqs.~(\ref{E*}) and (\ref{Edot}),
and the property $P_0 {\dot P}_0 P_0=0$,
we obtain 
\begin{eqnarray}
\mathrm{Tr}[P_0 (\partial_i H)P_0 (\partial_j H)]&=&\partial_iE_0\mathrm{Tr}[P_0 (\partial_j H)] \nonumber \\
&=& \frac{1}{g_0}\mathrm{Tr}[P_0 (\partial_i H)]\mathrm{Tr}[P_0 (\partial_j H)]
\label{app-id}
\end{eqnarray}
Substituting Eq.~(\ref{app-id}) into Eq.~(\ref{app-Gij}) and taking
the derivative with respect to $p$ yields
\begin{eqnarray}
\mathbf{G}_{ij}&=&\frac{1}{g_0}\int_{0}^{\infty} \mathrm{d}\tau ~\tau
e^{-p\tau}\Big(\mathrm{Tr}[P_0 (\partial_i H_{\tau}) (\partial_j H)] -
\frac{1}{g_{0}}\mathrm{Tr}[P_0 (\partial_i
  H)]\mathrm{Tr}[P_0(\partial_j H)] \Big)\Big |_{p=0},
\end{eqnarray} 
as desired.

\end{widetext}

\section{Detailed derivations of results reported in subsection~\protect{\ref{ISING-EXAMPLE}}}
\label{app:x(s)}

\subsection{Derivation of Eq.~(\ref{g-I-2})}
In the thermodynamic limit we replace $\sum_{\ell}$ by
$\frac{2m+1}{2\pi}\int_0^{\pi}\mathrm{d}\mathrm{z}$, where the
prefactor is due to a change of variables. Then Eq.~(\ref{p(x)}) yield
\begin{eqnarray}
p(\mathrm{x}) &=& \frac{1}{4}\sum_{\ell=1}^m \frac{\sin^2(\frac{2\pi
    \ell}{2m+1})}{[1-2(1+\cos \frac{2\pi
      \ell}{2m+1})(1-\mathrm{x})\mathrm{x}]^2} \notag \\
&\rightarrow& \int_0^{\pi}d\mathrm{z} \frac{\sin^2 \mathrm{z}}{[1-2(1+\cos
    \mathrm{z})(1-\mathrm{x})\mathrm{x}]^2}\notag \\
&=&\frac{\pi}{2(1-\mathrm{x})^2(1-2\mathrm{x})}.
\end{eqnarray}
Hence for $0\leq\mathrm{x'}<1/2$
\begin{eqnarray}
 \int_0^{\mathrm{x}(s)}\sqrt{p(\mathrm{x}^{\prime
 })}\mathrm{d}\mathrm{x}^{\prime }&=&
 \sqrt{\frac{\pi}{2}}\int_0^{\mathrm{x}(s)}\frac{\mathrm{d}\mathrm{x}^{\prime}}{(1-\mathrm{x}')
 \sqrt{1-2\mathrm{x}'}}\notag \\
&=&\frac{1}{2}\sqrt{\frac{\pi}{2}}[\pi-4\arctan\sqrt{1-2\mathrm{x}(s)}]\notag
 \\
 \int_0^{1/2}\sqrt{p(\mathrm{x}^{\prime
 })}\mathrm{d}\mathrm{x}^{\prime
 }&=&\frac{\pi}{2}\sqrt{\frac{\pi}{2}}. 
\end{eqnarray}
Now from Eq.~(\ref{105}) we obtain
\begin{eqnarray}
 2s &= & \int_0^{\mathrm{x}(s)}\sqrt{p(\mathrm{x}^{\prime
   })}\mathrm{d}\mathrm{x}^{\prime }/
 \int_0^{1/2}\sqrt{p(\mathrm{x}^{\prime
   })}\mathrm{d}\mathrm{x}^{\prime }\notag \\ 
& = & \frac{1}{\pi}[\pi-4\arctan\sqrt{1-2\mathrm{x}(s)}].
\end{eqnarray}
The last equation yields the first case in Eq.~(\ref{g-I-2}), 
\begin{eqnarray}
 \mathrm{x}(s)=\frac{1}{2}\left(1-\tan^2[\frac{\pi}{2}(s-\frac{1}{2})] \right).
\end{eqnarray}
The second case in Eq.~(\ref{g-I-2})) is obtained similarly.

%%%%%%%%%%%%%%%%%%%%%%%%%%%%%%%%%%%%%%%%%%%
\subsection{Derivation of Eq.~(\ref{g-I-bc})}

In the thermodynamic limit, using $\sum_{\ell} \rightarrow
\frac{2m+1}{2\pi}\int_0^{\pi}\mathrm{d}\mathrm{z}$ again,
\begin{eqnarray}
q(\mathrm{x})&=&\frac{1}{4}\sum_{\ell=1}^m\frac{\sin^2(\frac{2\pi
    \ell}{2m+1})}{[1-2\cos \frac{2\pi \ell}{2m+1}]^2} \notag \\
&\rightarrow&
\int_0^{\pi}d\mathrm{z}\frac{\sin^2 \mathrm{z}}{[1-2\mathrm{x}
    \cos\mathrm{z}+\mathrm{x}^2]^2}\notag \\ 
&=& \frac{\pi}{2(1-\mathrm{x}^2)}
\end{eqnarray}
Hence for $0\leq\mathrm{x}'\leq1$
\begin{eqnarray}
  \int_0^{\mathrm{x}(s)}\sqrt{q(\mathrm{x}^{\prime
  })}\mathrm{d}\mathrm{x}^{\prime }&=& \sqrt{\frac{\pi}{2}}\arcsin
  \mathrm{x}(s)\notag \\ 
  \int_0^{1}\sqrt{q(\mathrm{x}^{\prime })}\mathrm{d}\mathrm{x}^{\prime
  }&=& \frac{\pi}{2}\sqrt{\frac{\pi}{2}}. 
\end{eqnarray}
Now from Eq.~(\ref{108}) we obtain
\begin{eqnarray}
 s &= & \int_0^{\mathrm{x}(s)}\sqrt{q(\mathrm{x}^{\prime
   })}\mathrm{d}\mathrm{x}^{\prime }/
 \int_0^{1}\sqrt{q(\mathrm{x}^{\prime })}\mathrm{d}\mathrm{x}^{\prime
 }\notag \\ 
& = & \frac{2}{\pi}\arcsin\mathrm{x}(s). 
\end{eqnarray}
Thus we obtain Eq.~(\ref{g-I-bc}),
\begin{eqnarray}
 \mathrm{x}(s)=\sin(\pi s/2).
\end{eqnarray}

%%%%%%%%%%%%%%%%%%%%%%%%%%%%%%%%%%%%%%%%%%%

\subsection{Derivation of Eqs.~(\ref{ad-passage}) and (\ref{eq:chi})}

To solve
\begin{eqnarray}
 \ddot{\mathrm{X}}+ \nu\kappa \dot{\mathrm{X}}^2/2\mathrm{X}=0,
\end{eqnarray}
(where $\mathrm{X}\equiv \mathrm{x}-\mathrm{x}_c$) we use the following identity
\begin{eqnarray}
 \ddot{\mathrm{X}}=\dot{\mathrm{X}}\frac{\mathrm{d}\dot{\mathrm{X}}}{\mathrm{d}\mathrm{X}}=\frac{1}{2}\frac{\mathrm{d}}{\mathrm{d}\mathrm{X}}\left(
 \dot{\mathrm{X}}\right)^2.
\end{eqnarray}
 Hence
\begin{eqnarray}
 \frac{\mathrm{d}\dot{\mathrm{X}}^2}{\dot{\mathrm{X}}^2}&=&-\nu\kappa
 \frac{\mathrm{d}\mathrm{X}}{\mathrm{X}} \overset{\int}{\Rightarrow}
 \dot{\mathrm{X}}^2 =K \mathrm{X}^{-\nu\kappa} \Rightarrow
 \mathrm{X}^{\nu\kappa/2}\mathrm{d}\mathrm{X} = K \mathrm{d}s \notag \\
 &\overset{\int}{\Rightarrow}&
 \frac{\mathrm{X}^{\nu\kappa+1}}{\nu\kappa/2+1}=K(s-s_c)\notag \\
 &\Rightarrow& \mathrm{X}(s)=
      [K(\nu\kappa/2+1)(s-s_c)]^{\frac{1}{\nu\kappa/2+1}} \notag \\
&&~~~~~~~~\equiv
 A(s-s_c)^{\frac{2}{2+\nu\kappa}}. 
\end{eqnarray}
Therefore
\begin{eqnarray}
 \mathrm{x}(s)- \mathrm{x}_c=A(s-s_c)^{\frac{2}{2+\nu\kappa}}.
\end{eqnarray}

The derivation of Eq.~(\ref{eq:chi}) is:
\begin{eqnarray}
 \chi&=& \frac{2}{2+\nu\kappa}\notag \\
&\overset{\text{Eq.~(\ref{eq:kappa})}}{=}& \frac{2}{2+\nu(2\alpha-2z-d)}\notag \\
&\overset{\alpha=d+z-1/\nu}{ =} & \frac{2}{2+
   \nu(2d+2z-2/\nu-2z-d)}\notag \\
& =& \frac{2}{d\nu}.
\end{eqnarray}

%%\bibliographystyle{apsrev}
%%\bibliography{g-AQC-bib}

\begin{thebibliography}{57}
\bibitem[{g-A()}]{g-AQC-note}
\bibinfo{note}{{ We insert the prefactor $1/\sqrt{2g_0}$ into the definition in
  order to ensure $d(P_0,P'_0)\leq1$, because the maximum occurs when $P_0$ and
  $P'_0$ are orthogonal ($P_0P'_0=0$) }}.
\end{thebibliography}

%%%%%%%%%%%%%%%%%%%%%%%%%%%%%%%%%%%%%%%%%%%%%%%%%%%%%%%%%%%
\end{document}